\documentclass[11pt,notitlepage]{article}

\usepackage{latexsym,amssymb,epsfig,fullpage}
\usepackage{times}

\begin{document}
\newtheorem{defi}{Definition}
\newtheorem{theo}{Theorem}
\newtheorem{lem}[theo]{Lemma}
\newtheorem{cor}[theo]{Corollary}
\newtheorem{claim}[theo]{Claim}
\newtheorem{rem}{Remark}
\newtheorem{con}{Smoothness Assumption}
\newtheorem{prop}[theo]{Proposition}
\newtheorem{conjecture}{Conjecture}

\newcommand{\lemref}[1]{Lemma \ref{lem:#1}}

% Sam and Rachel

\newcommand{\Sa}{\mbox{$\cal A$}}
\newcommand{\mSa}{\cal S}
\newcommand{\Ra}{\mbox{$\cal B$}}
\newcommand{\mRa}{\cal R}
\newcommand{\SR}{\mbox{$\cal P$}}
\newcommand{\Si}{\mbox{$M$}}

%  For my thesis I use \varepsilon instead of \epsilon

\newcommand{\emptystr}[0]{\epsilon}
\newcommand{\eps}[0]{\varepsilon}
\newcommand{\com}[0]{\mbox{common}}
\renewcommand{\Pr}[0]{\mbox{\rm Prob}}
\renewcommand{\phi}{\varphi}
\newcommand{\Pre}[0]{\mbox{\bf{P}}}
\newcommand{\coin}[0]{\copyright\hspace{-2.1mm}
\raisebox{1pt}{$\scriptstyle |$}\hspace{1.5mm}}

%  Lambda calculus

\newcommand{\la}{\mbox{$\lambda$}}
\newcommand{\alc}{\mbox{$\mapsto_\alpha$}}
\newcommand{\bec}{\mbox{$\mapsto_\beta$}}
\newcommand{\becg}{\mbox{$\leadsto_\beta$}}
\newcommand{\becc}{\mbox{$\mapsto_\beta^*$}}
\newcommand{\beccg}{\mbox{$\leadsto_\beta^*$}}
\newcommand{\etc}{\mbox{$\mapsto_\eta$}}
\newcommand{\church}[1]{\mbox{$\overline{#1}$}}

% Complexity

\newcommand{\ds}{\mbox{{\sc Despace}}}
\newcommand{\ns}{\mbox{{\sc Nespace}}}
\newcommand{\dt}{\mbox{{\sc Dtemps}}}
\newcommand{\nt}{\mbox{{\sc Ntemps}}}

%  ADDING ``LOGLIKE FUNCTIONS''
\newcommand{\loglike}[1]{\mathop{#1}\nolimits}

%  DELIMITER PAIRS AND MATHEMATICAL FUNCTIONS

%\newcommand{\set}[1]{\left\{#1\right\}}
\newcommand{\enumset}[1]{\left\{#1\right\}}
\newcommand{\condset}[1]{\left\{\,#1\,\right\}}
\newcommand{\barset}[2]{\condset{#1\mid #2}}
\newcommand{\abs}[1]{\left|#1\right|}
\newcommand{\card}[1]{\left|#1\right|}
\newcommand{\bigo}[1]{O\left(#1\right)}
\newcommand{\Ocard}[1]{O\left(\card{#1}\right)}
\newcommand{\floor}[1]{\left\lfloor#1\right\rfloor}
\newcommand{\ceil}[1]{\left\lceil#1\right\rceil}
\newcommand{\ang}[1]{\ifmmode{\left\langle#1\right\rangle}
   \else{$\left\langle${#1}$\right\rangle$}\fi}
        % the \if allows use outside mathmode,
        % but will swallow following space there!
\newcommand{\prob}[1]{\Pr\left\{#1\right\}}
%\newcommand{\ds}[1]{\displaystyle{#1}}
%\newcommand{\iff}{\Leftrightarrow}

%  BINARY OPERATION, RELATION, AND ARROW SYMBOLS

\newcommand{\propsubset}
    {\stackrel{\subset}{\scriptscriptstyle \not {\scriptstyle =}}}
\newcommand{\propsubs}{\stackrel{.}{-}}
\newcommand{\polyred}{\leq_p}
\newcommand{\logred}{\leq_L}
\newcommand{\dhrightarrow}{\rightarrow \hspace*{-15pt} \rightarrow}
\newcommand{\isin}{\stackrel{\scriptscriptstyle ?}{\in}}
\newcommand{\iseq}{\stackrel{\scriptscriptstyle ?}{=}}

%  OTHER MATHEMATICAL SYMBOLS OR CONSTRUCTS

\newcommand{\half}{\frac{1}{2}}
\newcommand{\mr}{\rhd}
\newcommand{\ml}{\lhd}
\newcommand{\nm}{\bigtriangledown}
\newcommand{\is}{\leftarrow}
\newcommand{\fto}{\rightarrow}
\newcommand{\inr}{\in_{\mbox{R}}}
\newcommand{\non}[1]{\bar{#1}}
\renewcommand{\non}[1]{\bar{#1}}
\newcommand{\una}[1]{1^{#1};}
\newcommand{\lgn}{\lg n}
\newcommand{\degrees}{^{\circ}}
\newcommand{\implies}{\Rightarrow}
\renewcommand{\choose}[2]{{{#1}\atopwithdelims(){#2}}}
\newcommand{\OTchoose}[2]{{(\mbox{}_{#2}^{#1})}}
\newcommand{\js}[2]{\left(\frac{#1}{#2}\right)}  % Jacobi symbol
\newcommand{\ls}[2]{\choose{#1}{#2}}  % Legendre symbol
\newcommand{\lap}[1]{{\cal L}(#1)}  % Laplace Transform
\newcommand{\lapinv}[1]{{\cal L}^{-1}(#1)}  % Inverse Laplace Transform
\newcommand{\sumd}[2]{\displaystyle \sum\limits_{#1}^{#2}}
\newcommand{\prodd}[2]{\displaystyle \prod\limits_{#1}^{#2}}
\newcommand{\xor}[2]{\displaystyle \bigoplus\limits_{#1}^{#2}}
\newcommand{\uni}[2]{\displaystyle \bigcup\limits_{#1}^{#2}}
\newcommand{\maj}[2]{\displaystyle\mathrel{\mathop{\mbox{\rm maj}}\limits_{#1}^{#2}}}
\newcommand{\DO}[2]{\displaystyle\mathrel{\mathop{\mbox{\bf DO}}\limits_{#1}^{#2}}}

\newcommand{\goesto}[1]{\stackrel{#1}{\rightarrow}}
\newcommand{\pgoesto}[1]{\stackrel{#1}{\dhrightarrow}}
\newcommand{\loops}{\uparrow}
\newcommand{\stops}{\downarrow}

%\renewcommand{\prop}[2]{#1 \mbox{  } \left[ #2 \right]}

%  VECTORS AND MATRICES

\newcommand{\seq}[3]{#1_{#2} \cdots #1_{#3}}
\newcommand{\mat}[5]
    {\left(
        \begin{array}{ccc}
        #1_{#2,#4} & \ldots & #1_{#2,#5}\\
        \vdots & \ddots & \vdots\\
        #1_{#3,#4} & \ldots & #1_{#3,#5}
        \end{array}
     \right)}
\newcommand{\matl}[6]
    {\left(
        \begin{array}{cccc}
        #1_{#2,#4} & \ldots & \ldots & #1_{#2,#5}\\
        \vdots & & & \vdots\\
        #1_{#6-1,#4} & \ldots & \ldots &  #1_{#6-1,#5}\\
        #1_{#6+1,#4} & \ldots & \ldots &  #1_{#6+1,#5}\\
        \vdots & & & \vdots\\
        #1_{#3,#4} & \ldots & \ldots & #1_{#3,#5}
        \end{array}
     \right)}
\newcommand{\vect}[4]
    {\parbox{#4}
        {$#1_{#2}$\\
         \makebox[#4]{.}\\
         \makebox[#4]{.}\\
         \makebox[#4]{.}\\
         $#1_{#3}$}}

%  CRYPTOGRAPHY

\newcommand{\bc}[1]{\mbox{\fbox{$#1$}}}

\newcommand{\eBOT} {Oblivious Bit Transfer } 
\newcommand{\eSOT} {Oblivious String Transfer } 
\newcommand{\eQOT} {Oblivious Qit Transfer } 

\newcommand{\QOT}[2] {$\mbox{OT}^{ #2 }$} 
\newcommand{\GQOT}[2] {$\mbox{GOT}^{ #2 }$}
\newcommand{\UQOT}[2] {$\mbox{UOT}^{ #2 }$}
\newcommand{\TOQ}[2] {$\mbox{}^{ #2 }\mbox{TO}$} 
\newcommand{\BOT} {\QOT{}{}} 
\newcommand{\GBOT} {\GQOT{}{}}
\newcommand{\UBOT} {\UQOT{}{}} 
\newcommand{\TOB} {\TOQ{}{}} 
\newcommand{\SOT} {OST} 
\newcommand{\QT} {QT}

\newcommand{\obt}[2] {\mbox{$\OTchoose{ #2 }{ #1 }$--\BOT}}
\newcommand{\gobt}[2] {\mbox{$\OTchoose{ #2 }{ #1 }$--\GBOT}}
\newcommand{\uobt}[2] {\mbox{$\OTchoose{ #2 }{ #1 }$--\UBOT}}
\newcommand{\tbo}[2] {\mbox{\TOB--$\OTchoose{ #2 }{ #1 }$}}
\newcommand{\oqt}[4] {\mbox{$\OTchoose{ #2 }{ #1 }$--\QOT{ #3 }{ #4 }}}
\newcommand{\tqo}[4] {\TOQ{ #3 }{ #4}--\mbox{$\OTchoose{ #2 }{ #1 }$}}
\newcommand{\sot}[2] {\mbox{$\OTchoose{ #2 }{ #1 }$--\SOT}}

\newcommand{\eot}[3] {\mbox{$\OTchoose{ #3 }{ #2 } #1$--\BOT}}

\newcommand{\ots} {\obt{1}{2}}
\newcommand{\gots} {\gobt{1}{2}}
\newcommand{\uots} {\uobt{1}{2}}
\newcommand{\sto} {\tbo{1}{2}}
\newcommand{\otq}[2] {\oqt{1}{2}{ #1 }{ #2 }}
\newcommand{\qto}[2] {\tqo{1}{2}{ #1 }{ #2 }}
\newcommand{\xtq} {\mbox{$\OTchoose{2}{1}$--$\mbox{XOT}$}}

\newcommand{\eots} {\eot{\alpha}{1}{2}}

\newcommand{\ntr}[1] {\mbox{$ #1 $--NT}}

%%TEXTURES PICTURES

\def\picture #1 by #2 (#3){
  \vbox to #2{
    \hrule width #1 height 0pt depth 0pt
    \vfill
    \special{picture #3} % this is the low-level interface
    }
  }

\def\scaledpicture #1 by #2 (#3 scaled #4){{
  \dimen0=#1 \dimen1=#2
  \divide\dimen0 by 1000 \multiply\dimen0 by #4
  \divide\dimen1 by 1000 \multiply\dimen1 by #4
  \picture \dimen0 by \dimen1 (#3 scaled #4)}
  }

\newcommand{\FI}{{\cal F}}
%G J'ai change la police pour le F de Field
\newcommand{\nats}{{\cal N}}
%G j'ai ajoute \nats pour le N des naturels
\newcommand{\FAM}{{\cal F}am}

\newcommand{\blackslug}{\hbox{\hskip 1pt \vrule width 4pt height 8pt
depth 1.5pt \hskip 1pt}}
\newcommand{\QED}{\quad\blackslug\lower 8.5pt\null}

\newcommand{\inverse}[1]{\frac{1}{{#1}}}       % 1/x, inverse
\newcommand{\inv}[1]{{#1}^{-1}}                % x^{-1}, inverse

\newcommand{\zeromat}[0]{
{\tiny
\begin{array}{cccc}
0 & 0 & \dots & 0\\
\vdots & \vdots & \ddots & \vdots\\
0 & 0 & \dots & 0
\end{array}
}
}

\newcommand{\idmat}[0]{
{\tiny
\begin{array}{cccc}
1 & 0 & \dots & 0\\
0 & 1 & \dots & 0\\
\vdots & \vdots & \ddots & \vdots\\
0 & 0 & \dots & 1
\end{array}
}
}

\newcommand{\idmatx}[0]{
{\tiny
\begin{array}{ccccc}
1 & 0 & \dots & 0 & 0\\
0 & 1 & \dots & 0 & 0\\
\vdots & \vdots & \ddots & \vdots & \vdots\\
0 & 0 & \dots & 1 & 0
\end{array}
}
}

%  MACROS FOR ALGORITHMS OR PROGRAMS     REQUIRES genmacs.tex

\newcounter{proglevel}
\newcounter{progline}
\newlength{\linenosep}
\newlength{\subprogindent}
\newlength{\progitemhangindent}
\newlength{\addprogitemsep}

\newenvironment{program}[1]%
{\setcounter{proglevel}{0}\setcounter{progline}{0}
 \settowidth{\linenosep}{\quad}
 \settowidth{\subprogindent}{\qquad}\settowidth{\progitemhangindent}{\quad}
 \setlength{\addprogitemsep}{0pt}
\newcommand{\setsubprogtabs}{#1\=\hspace*{\linenosep}
                             \hspace*{\value{proglevel}\subprogindent}\=
                             \hspace*{\progitemhangindent}\=\+\+\+\kill}
 \begin{tabpackage}\setlength{\tabbingsep}{0pt}\setsubprogtabs}%
{\end{tabpackage}}

% NOT TO BE USED AS AN ENVIRONMENT; USE ONLY \subprogram AND \endsubprogram
\newenvironment{subprogram}%
{\stepcounter{proglevel}\pushtabs\<\<\<\-\-\-\setsubprogtabs}%
{\poptabs\addtocounter{proglevel}{-1}}

\newif\ifprogitemarg
\makeatletter
\def\progitem{\@ifnextchar [{\progitemargtrue\progitemwarg}%
{\progitemargfalse\progitemwarg[\proglinenum]}}
\makeatother

\newlength{\progitemstrutheight}
\setlength{\progitemstrutheight}{0pt} % really start at {\baselineskip} ??
\addtolength{\progitemstrutheight}{\addprogitemsep}

\def\progitemwarg[#1]{\<\< #1 \'\>\rule{0pt}{\progitemstrutheight}%
\ifprogitemarg\else%
\addtocounter{progline}{-1}\refstepcounter{progline}\ignorespaces\fi}

\newcommand{\proglinenum}{\refstepcounter{progline}\theprogline}

\newcommand{\proglabel}[1]{\label{#1}\ignorespaces}

\newcommand{\progcom}[1]{[{\it #1}]}

\newcommand{\ifthenelse}[3]     % OLD COMMAND; MAY BE BOGUS IN CURRENT LATEX
    {{\it if} #1 \begin{description}
                                  \item[\hfill {\it then}] #2
                                  \item[\hfill {\it else}] #3
                              \end{description}}

\newcommand{\Begin}{{\bf begin\ }}
\newcommand{\End}{{\bf end\ }}
\newcommand{\Do}{{\bf do\ }}
\newcommand{\Od}{{\bf od}}
\newcommand{\Enddo}{{\bf enddo}}
\newcommand{\Loop}{{\bf loop\ }}
\newcommand{\Endloop}{{\bf endloop}}
\newcommand{\For}{{\bf for\ }}
\newcommand{\Endfor}{{\bf endfor}}
\newcommand{\Foreach}{{\bf foreach\ }}
\newcommand{\To}{{\bf to\ }}
\newcommand{\If}{{\bf if\ }}
\newcommand{\Then}{{\bf then\ }}
\newcommand{\Else}{{\bf else\ }}
\newcommand{\Elif}{{\bf elif\ }}
\newcommand{\Fi}{{\bf fi}}
\newcommand{\Endif}{{\bf endif}}
\newcommand{\While}{{\bf while\ }}
\newcommand{\Endwhile}{{\bf endwhile}}
\newcommand{\Repeat}{{\bf repeat\ }}
\newcommand{\Until}{{\bf until\ }}
\newcommand{\Case}{{\bf case\ }}
\newcommand{\Of}{{\bf of\ }}
\newcommand{\Procedure}{{\bf procedure\ }}
\newcommand{\Call}{{\bf call\ }}
\newcommand{\With}{{\bf with\ }}
\newcommand{\Imports}{{\bf imports\ }}
\newcommand{\Var}{{\bf var\ }}
\newcommand{\Type}{{\bf type\ }}
\newcommand{\Assert}{{\bf assert\ }}
\newcommand{\True}{{\rm true}}
\newcommand{\False}{{\rm false}}

% PROTOCOLS and ALGOs

\newcounter{protoline}
\newlength{\boxwidth}
\newlength{\bigboxwidth}

\setlength{\boxwidth}{\textwidth}
\addtolength{\boxwidth}{-0.6cm}
\setlength{\bigboxwidth}{\textwidth}
\addtolength{\bigboxwidth}{-0.1cm}

\newtheorem{proto_body}{Protocol}[section]
\newtheorem{algo_body}{Algorithm}[section]

\newenvironment{protocol}[1]{ 
\setcounter{protoline}{0}
\begin{minipage}{\boxwidth}
\newcommand{\step}[1]
        {\addtocounter{protoline}{1} {\item [\ \ \ \ \arabic{protoline}:]{\small##1 } }}
\begin{proto_body}[ #1 ]
\ \\[-5mm] 
\begin{description} }{\end{description}\end{proto_body}\end{minipage}}
\newenvironment{algorithm}[1]{ 
\setcounter{protoline}{0}
\begin{minipage}{\boxwidth}
\newcommand{\step}[1]
        {\addtocounter{protoline}{1} {\item [\ \ \ \ \arabic{protoline}:]{\small##1 } }}
\begin{algo_body}[ #1 ]
\ \\ 
\begin{description} }{\end{description}\end{algo_body}\end{minipage}}

\newcommand{\proto}[2]{\ \\ \fbox{ \begin{protocol}{#1}#2 \end{protocol}}\ \\}
\newcommand{\algo}[2]{\ \\ \fbox{ \begin{algorithm}{#1}#2 \end{algorithm}}\ \\}

\newenvironment{protocol_cont}[0]{ 
\begin{minipage}{\boxwidth}
\addtocounter{proto_body}{-1}
\newcommand{\step}[1]
        {\addtocounter{protoline}{1} {\item [\ \ \ \ \arabic{protoline}:]{\small##1 } }}
\begin{proto_body}
\ \\ 
\begin{description} }{\end{description}\end{proto_body}\end{minipage}}

\newcommand{\cont}[1]{\ \\ \fbox{ \begin{protocol_cont}#1 \end{protocol_cont}}\ \\}

\newcommand{\dist}[2]{
\newcommand{\alt}[2] {##1 & \mbox{\rm : } ##2}
$ #1 \inr \left\{
\begin{array}{ll}
#2
\end{array}
\right.$
}

% REDUCTION

\newcounter{reduline}
\newlength{\boxwidthr}

\setlength{\boxwidthr}{\textwidth}
\addtolength{\boxwidthr}{-0.6cm}

\newtheorem{redu_body}{Reduction}[section]
\newenvironment{reduction}[2]{ 
\setcounter{reduline}{0}
\begin{minipage}{\boxwidthr}
\newcommand{\step}[1]
        {\addtocounter{reduline}{1} {\item [\ \ \ \ \arabic{reduline}:]{\small##1 } }}
\begin{redu_body}[ #1 from #2 ]
\ \\ 
\begin{description} }{\end{description}\end{redu_body}\end{minipage}}

\newcommand{\redu}[3]{\ \\ \fbox{ \begin{reduction}{#1}{#2}#3 \end{reduction}}\ \\}

\newenvironment{redu_cont}[0]{ 
\begin{minipage}{\boxwidthr}
\addtocounter{redu_body}{-1}
\newcommand{\step}[1]
        {\addtocounter{reduline}{1} {\item [\ \ \ \ \arabic{reduline}:]{\small##1 } }}
\begin{redu_body}
\ \\ 
\begin{description} }{\end{description}\end{redu_body}\end{minipage}}

\newcommand{\rcont}[1]{\ \\ \fbox{ \begin{redu_cont}#1 \end{redu_cont}}\ \\}

\newcommand{\probab}{{\rm Prob\, }}
\newcommand{\CE}{{\cal E}}
\newcommand{\OCE}{\overline{{\cal E}}}
\newcommand{\lt}{\leadsto}
\newcommand{\varl}{l}
\newcommand{\ul}{\underline}
\newcommand{\noi}{\noindent}
\newcommand{\cancel}[1]{}
\newcommand{\canpro}[1]{}
\newcommand{\vs}[1]{\vspace{#1mm}}
\newcommand{\vsk}{\vspace{3mm}}
\newcommand{\vsl}{\vspace{7mm}}
\newcommand{\ind}{\rule{7mm}{0mm}}
\newcommand{\rh}[1]{\rule{#1 mm}{0mm}}
\newcommand{\rv}[1]{\rule{0mm}{#1 mm}}
\newcommand{\rz}{\rule{0mm}{0mm}}
\newcommand{\be}{\beta}
\newcommand{\beq}{\begin{equation}}
\newcommand{\ee}{\end{equation}}
\newcommand{\eeq}{\end{equation}}
\newcommand{\bea}{\begin{eqnarray}}
\newcommand{\eea}{\end{eqnarray}}
\newcommand{\beas}{\begin{eqnarray*}}
\newcommand{\eeas}{\end{eqnarray*}}
\newcommand{\bece}{\begin{center}}
\newcommand{\ence}{\end{center}}
\newcommand{\beit}{\begin{itemize}}
\newcommand{\enit}{\end{itemize}}
\newcommand{\nli}{\newline}
\newcommand{\hyp}{\hyphenation}
\newcommand{\mod}{{\rm mod\ }}
\newcommand{\co}{{\cal O}}
\newcommand{\ce}{{\cal E}}
\newcommand{\cu}{{\cal U}}
\newcommand{\oce}{\overline{{\cal E}}}
\newcommand{\lgr}{\langle g\rangle}
\newcommand{\proofstart}{{\it Proof: }}
\newcommand{\proofsketchstart}{{\it Proof (sketch): }}
\newcommand{\proofend}{\hspace*{\fill} $\Box$\\}
\newcommand{\pe}{\hspace*{\fill} $\Box$\\}
\newcommand{\DHO}{{\rm DHO}}
\newcommand{\yes}{{\tt yes}}
\newcommand{\no}{{\tt no}}
\newcommand{\dhd}{{\rm DHD}}
\newcommand{\RSA}{{\rm RSA}}
\newcommand{\ord}{{\rm ord}}
\newcommand{\OC}{{\rm OC}}
\newcommand{\OCs}{{\rm OCs}}
\newcommand{\fp}{GF(p)}
\newcommand{\fq}{{\bf F}_q}
\newcommand{\zn}{{\bf Z}_n}
\newcommand{\znn}{{\bf Z}/{n{\bf Z}}}
\newcommand{\z}{{\bf Z}}
\newcommand{\zpt}{{\bf Z}_{p^t}}
\newcommand{\zpe}{{\bf Z}_{p^e}}
\newcommand{\zz}{{\bf Z}}
\newcommand{\zznn}{{\bf Z}_n}
\newcommand{\al}{\alpha}
\newcommand{\Chi}{\chi}
\newcommand{\fqN}{{\bf F}_{q^N}}
\newcommand{\fpn}{{\bf F}_{p^n}}
\newcommand{\fpk}{{\bf F}_{p^k}}
\newcommand{\fpi}{{\bf F}_{p^i}}
\newcommand{\fpg}{{\bf F}_{p^g}}
\newcommand{\fpm}{{\bf F}_{p^m}}
\newcommand{\fqn}{{\bf F}_{q^n}}
\newcommand{\fqi}{{\bf F}_{q^i}}
\newcommand{\fpN}{{\bf F}_{p^N}}
\newcommand{\ep}{\varepsilon}
\newcommand{\de}{\delta}
\newcommand{\get}{\tilde{g}}
\newcommand{\vp}{\varphi}
\newcommand{\hterm}[1]{\mbox{{\rm hterm(}$#1${\rm )}}}
\newcommand{\MM}[1]{\mbox{{\rm M(}$#1${\rm )}}}
\newcommand{\hcoeff}[1]{\mbox{{\rm hcoeff(}$#1${\rm )}}}
\newcommand{\fei}[2]{\mbox{{\rm #1(}$#1${\rm )}}}
\newcommand{\DD}{{\bf D}}
\newcommand{\PP}{{\bf P}}
\newcommand{\JJ}{{\bf J}}
\newcommand{\di}[2]{\mbox{{\rm div(}$#1,#2${\rm )}}}
\newcommand{\spoly}[2]{\mbox{{\rm s-poly(}$#1,#2${\rm )}}}
\newcommand{\lcm}[2]{\mbox{{\rm lcm(}$#1,#2${\rm )}}}
\newcommand{\fzw}[3]{\mbox{{\rm #1(}$#2,#3${\rm )}}}
\newcommand{\gcdd}[2]{\mbox{{\rm gcd(}$#1,#2${\rm )}}}
\newcommand{\OO}{{\bf 0}}
\newcommand{\Op}{{\bf 0}_p}
\newcommand{\Oq}{{\bf 0}_q}
\newcommand{\ga}{\gamma}
\newcommand{\si}{\sigma}
\newcommand{\ra}{\rightarrow}
\newcommand{\lra}{\longrightarrow}
\newcommand{\Lra}{\Longrightarrow}
\newcommand{\llra}{\longleftrightarrow}
\newcommand{\Llra}{\Longleftrightarrow}
\newcommand{\toinf}{\rightarrow\infty}
\newcommand{\eqdef}{\stackrel{{\scriptstyle\bigtriangleup}}{=}}
\newcommand{\ff}{{\bf F}}
\newcommand{\dhh}{{\rm DH}}
\newcommand{\pp}{{\rm P}}
\newcommand{\gd}{\Delta}
\newcommand{\next}{}

\newcommand{\zeotks}{\choose{2}{1}\mbox{{\rm -OT}}^k_s}
\newcommand{\zeot}{\choose{2}{1}\mbox{{\rm -OT}}}

%\includeonly{goal}

\newcommand{\huh}[1]{{\bf **#1**}}
\hyphenation{pre-sents pre-sent}
\newcommand{\HH}{\mbox{\bf H}}
\newcommand{\entr}[1]{\mbox{\bf H}\left(#1\right)}
\newcommand{\info}[1]{\mbox{\bf I}\left(#1\right)}
\newcommand{\BSa}{\mbox{$\tilde {\cal A}$}}
\newcommand{\BRa}{\mbox{$\tilde {\cal B}$}}
\newcommand{\HSa}{\mbox{$\bar {\cal A}$}}
\newcommand{\HRa}{\mbox{$\bar {\cal B}$}}

\newcommand{\bigbox}[1]{\fbox{\begin{minipage}{\textwidth}#1\end{minipage}}}

\newlength{\protlr}\newlength{\protm}\newlength{\protarr}
\newlength{\protraise}
\setlength{\protlr}{3.5cm}
\setlength{\protm}{2cm}
\setlength{\protarr}{\protm}\addtolength{\protarr}{-.1cm}
\setlength{\protraise}{1.5ex}

%                    Patentes Quantinque

\newcommand{\breidbart}{{\cal B}}

\newcommand{\bwedge}{\!\bigwedge\!}
\newcommand{\bvee}{\!\bigvee\!}

\newcommand{\bra}[1]{\langle #1 |}
\newcommand{\ket}[1]{| #1 \rangle}
\newcommand{\braket}[2]{\langle #1 | #2 \rangle}
\newcommand{\kett}[2]{\braket{#1}{#2}}

\newcommand{\vpol}{\uparrow}
\newcommand{\vpols}{\mbox{$\updownarrow$}\:}
\newcommand{\hpol}{\rightarrow}
\newcommand{\hpols}{\mbox{$\leftrightarrow$}}
\newcommand{\rbas}{+}
\newcommand{\rbass}{\mbox{$\hpols$}\llap{\mbox{$\vpols$}}}
\newcommand{\dbas}{\times}
\newcommand{\dbass}{\mbox{$\ppols$}\llap{\mbox{$\qpols$}}}
\newcommand{\cbas}{\mbox{\footnotesize $\bigcirc$}}
\newcommand{\cbass}{\mbox{\scriptsize $\bigcirc$}}
\newcommand{\chck}{\mbox{$\surd$}}

\newcommand{\ppol}{\nearrow}
\newcommand{\qpol}{\nwarrow}
\newcommand{\ppols}{\scriptsize
\,\mbox{$\nearrow$}\llap{\mbox{$\swarrow$}}\,}
\newcommand{\qpols}{\scriptsize
\,\mbox{$\searrow$}\llap{\mbox{$\nwarrow$}}\,}

\newcommand{\comment}[1]{}

\newcommand{\set}[2]%
{ \left\{\kern-0.4em
    \begin{array}{l|r}
      #1 & #2
    \end{array}
  \kern-0.4em\right\}
}

% right arrow with text #1
\newcommand{\rarrow}[1]%
{ \setlength{\unitlength}{\protarr}%
  \raisebox{0.5ex}%%
  { \begin{picture}(0,0)%
      \thinlines%
      \put(0,0){\vector(1,0){1}}%
    \end{picture}%
  }%
  \raisebox{\protraise}{\makebox[\protarr]{#1}}%
}

% left arrow with text #1
\newcommand{\larrow}[1]%
{ \setlength{\unitlength}{\protarr}%
  \raisebox{0.5ex}%%
  { \begin{picture}(0,0)%
      \thinlines%
      \put(1,0){\vector(-1,0){1}}%
    \end{picture}%
  }%
  \raisebox{1.0ex}{\makebox[\protarr]{#1}}%
}

%%%%%%%%%%%%%%%%%%%%%%%%%%%%%%%%%%%%%%%%%%%%%%%%%%%%%%%%%%%%%%%%%%%%%%%%%%%%%
%%%%%%%%%%%%%%%%%%%%%%%%%%%%%%%%%%%%%%%%%%%%%%%%%%%%%%%%%%%%%%%%%%%%%%%%%%%%%
%%%%%%%%%%%%%%%%%%%%%%%%%%%%%%%%%%%%%%%%%%%%%%%%%%%%%%%%%%%%%%%%%%%%%%%%%%%%%

\newcommand{\hfsq}{{\frac {1} {\sqrt{2}}}}
\newcommand{\Tr}{{\mathrm{Tr}}}

\newcommand{\an}[1]{}

\newenvironment{proofof}[1]{\noindent \emph{Proof (of #1):}}{\proofend}
\newenvironment{proof}{\noindent \emph{Proof:}}{\proofend}
\newenvironment{sketch}{\noindent \emph{Proof (sketch):}}{\proofend}

\newcommand{\paren}[1]{\left( {#1} \right)}
\renewcommand{\set}[1]{\left\{ {#1} \right\}}
\newcommand{\qas}{\textsc{qas}}
\newcommand{\proj}[1]{\ket{#1}\bra{#1}}

\newcommand{\ze}{\mathbb{Z}}

\newcommand{\acc}{\textsc{acc}}
\newcommand{\rej}{\textsc{rej}}

\newcommand{\A}{\Sa}
\newcommand{\B}{\Ra}
\newcommand{\C}{${\cal C}$}
\newcommand{\Ca}{C_1}
\newcommand{\Cb}{C_2}
\newcommand{\Caperp}{C_1^\perp}

\newcommand{\epr}[1]{\ket{\Phi^+}^{\otimes {#1}}}

\title{Authentication of Quantum Messages}
      
\author{% 
Howard Barnum\,% 
\thanks{\, CCS-3 Group, Los Alamos National Laboratories, Los Alamos,
New Mexico 87554 USA.  e-mail: {\tt barnum@lanl.gov}.  Supported by 
the US DOE; part of this research
was done while working at the University of Bristol (UK), and supported
by the EU QAIP Consortium (IST-1999-11234).}
,\ Claude Cr\'epeau\,%
\thanks{\, School of Computer Science, McGill University,
Montr\'eal (Qu\'ebec), Canada. e-mail: {\tt crepeau@cs.mcgill.ca}.
\,Supported in part by Qu\'ebec's FCAR and Canada's NSERC.}
,\ Daniel Gottesman\,%
\thanks{\, UC Berkeley, EECS: Computer Science Division,
Soda Hall 585, Berkeley, California 94720, USA. e-mail:
{\tt gottesma@eecs.berkeley.edu}.
\,Supported by the Clay Mathematics Institute.}
,\ Adam Smith\,%
\thanks{\, M.I.T., Laboratory for Computer Science,
200 Technology Square, Cambridge MA 02139,
USA. e-mail: {\tt asmith@theory.lcs.mit.edu}.
Supported in part by U.S. Army Research Office Grant
    DAAD19-00-1-0177. Some of this research was done while the author
    was visiting McGill University.}
, and\,  Alain Tapp
\thanks{\,D\'epartement IRO, Universit\'e de Montr\'eal,
C.P. 6128, succursale centre-ville,
Montr\'eal (Qu\'ebec), Canada H3C 3J7.  e-mail: {\tt tappa@iro.umontreal.ca}.
Part of this research was done while working at Department of Combinatorics and Optimization,  University of Waterloo and McGill University.}
%\thanks{\, Centre for Applied Cryptographic Research (CACR), 
%Department of Combinatorics and Optimization,  University of 
%Waterloo, Waterloo, Canada. e-mail: {\tt atapp@cacr.math.uaterloo.ca}.
% Supported by Canada's NSERC.}
}

\date{}

\maketitle

\thispagestyle{empty}

\begin{abstract}
  \noindent
  Authentication is a well-studied area of classical cryptography: a
  sender \Sa\ and a receiver \Ra\ sharing a classical private key want
  to exchange a classical message with the guarantee that the message
  has not been modified or replaced by a dishonest party with
  control of the communication line. In this paper we study the
  authentication of messages composed of {\em quantum states}.
  
  We give a formal definition of authentication in the quantum
  setting.  Assuming \Sa\ and \Ra\ have access to an insecure quantum
  channel and share a private, classical random key, we provide a {\em
    non-interactive} scheme that both enables \Sa\ to encrypt and
  authenticate (with unconditional security) an $m$ qubit message by
  encoding it into $m+s$ qubits, where the probability decreases
  exponentially in the security parameter $s$.  The scheme requires a
  private key of size $2m+O(s)$. To achieve this, we give a highly
  efficient protocol for testing the purity of shared EPR pairs. 

  It has long been known that learning information about a general
  quantum state will necessarily disturb it. We refine this result to
  show that such a disturbance can be done with few side effects,
  allowing it to circumvent cryptographic protections.  Consequently,
  any scheme to authenticate quantum messages must also encrypt them.
  In contrast, no such constraint exists classically: authentication
  and encryption are independent tasks, and one can authenticate a
  message while leaving it publicly readable.
%
%   If you could learn any informatieon, not only would you (necessarily)
%   disturb the state, but you can alter the state without being
%   detected.
%
  
  This reasoning has two important consequences: On one hand, it
  allows us to give a lower bound of $2m$ key bits for authenticating
  $m$ qubits, which makes our protocol asymptotically optimal. On the
  other hand, we use it to show that digitally signing
  quantum states is impossible, even with only computational security.

\end{abstract}

% \newpage
% \thispagestyle{empty}
% {\small \tableofcontents}

\newpage
\pagenumbering{arabic}

\section{Introduction}\label{secintro}

\newcommand{\mypar}[1]{\vskip 6pt \noindent  {\bf #1}\ }

Until recently, the expression ``quantum cryptography'' referred
mostly to quantum key distribution protocols \cite{BB84,B92,E90}.
However, these words now refer to a larger set of
problems. % involving both classical and quantum data.  
While QKD and many other quantum protocols attempt to provide improved
security for tasks involving classical information, an emerging area
of quantum cryptography attempts instead to create secure protocols
for tasks involving {\em quantum} information.  One standard
cryptographic task is the {\em authentication} of messages: \Sa\ 
transmits some information to \Ra\ over an insecure channel, and they
wish to be sure that it has not been tampered with {\it en route}.
When the message is classical, and \Sa\ and \Ra\ share a random
private key, this problem can be solved by, for instance, the
Wegman-Carter scheme \cite{carweg79}.  In this paper, we discuss the
analogous question for quantum messages.

\mypar{A naive approach} 
If we assume \Sa\ and \Ra\ share a private {\em quantum} key in the
form of $m$ EPR pairs, as well as some private classical key, there is
a straightforward solution to this problem: \Sa\ simply uses quantum
teleportation \cite{BBCJPW00} to send her message to Bob,
authenticating the $2m$ classical bits transmitted in the
teleportation protocol.  If \Sa\ and \Ra\ initially share only a
classical key, however, the task is more difficult.  We start with a
simple approach: first distribute EPR pairs (which might get corrupted
in transit), and then use entanglement purification \cite{BBPSSW96} to
establish clean pairs for teleportation.  This can be improved: we do
not need a full-scale entanglement purification protocol, which
produces good EPR pairs even if the channel is noisy; instead we only
need something we call a {\em purity testing protocol}, which checks
that EPR pairs are correct, but does not attempt to repair them in
case of error.
% The security of an authentication protocol involving
% purity testing followed by teleportation can be proven with methods
% originally used by Lo and Chau \cite{LC99} to show the security of key
% distribution.

Unfortunately, any such protocol will have to be interactive, since
\Sa\ must first send some qubits to \Ra and then wait for
confirmation of receipt before completing the transmission.  This is
unsuitable for situations in which a message is stored and must be
checked for authenticity at a later time.  Also, this interactive
protocol achieves something stronger than what is required of a
quantum authentication scheme: at the end of the purity-testing based
scheme, {\em both} Alice and Bob know that the transmission was
successful, whereas for authentication, we only require that Bob
knows.

\mypar{Contributions}
In this paper we study \emph{non-interactive} quantum authentication
schemes with classical keys. Our primary contributions are:

\begin{description}
\item[{\rm $\bullet$ Formal definition of authentication for quantum
    states}]{~}
  
  In classical authentication, one simply limits the probability that
  the adversary can make \emph{any} change to the state without
  detection. This condition is too stringent for quantum information,
  where we only require high fidelity to the original state. We state
  our definition in terms of the transmission of pure states
  (section~\ref{sec:qa}), but also show that the same definition
  implies security for mixed or entangled states.
  
\item[{\rm $\bullet$ Construction of efficient purity testing protocols
%    based on projective geometry and quantum stabilizer coding
  }]~
  
  We show how to create purity-testing protocols using families of
  quantum error-correcting codes with a particular covering property,
  namely that any Pauli error is detected by most of the codes in the
  family.  We construct an efficient such family based on projective
  geometry, yielding a purity-testing protocol requiring only $O(s)$
  (classical) bits of communication, where $s$ is the security
  parameter (section~\ref{sec:ptc}).  
  
  Purity-testing codes have not explicitly appeared before in the
  literature, but have been present implicitly in earlier work, for
  instance \cite{LC99,SP00}.  To prove our purity-testing protocols
  secure, we use a ``quantum-to-classical'' reduction, due to Lo and
  Chau \cite{LC99}.  Subsequently to our work, Ambainis, Smith, and
  Yang \cite{ASY} used our construction of purity-testing protocols in
  a study of more general entanglement extraction procedures.

\item[{\rm $\bullet$ Construction of non-interactive quantum
    authentication schemes (\qas)}]{~}
  
  We show that a secure non-interactive \qas\ can be constructed from
  any purity-testing protocol derived, as above, from {\sc qecc}s
  (section~\ref{sec:prot}). In particular, for our family of codes, we
  obtain an authentication scheme which requires sending $m+s$ qubits,
  and consuming $2m+O(s)$ bits of classical key for a message of $m$
  qubits.  The proof techniques in the Shor and Preskill paper
  \cite{SP00} serve as inspiration for the transformation from an
  interactive purity-testing protocol to a non-interactive
  \qas.

\item[{\rm $\bullet$ Study of the relation between encryption and
    authentication}]{~}

  One feature of our authentication protocol is that it completely
  encrypts the quantum message being sent.  We show that this is a
  necessary feature of {\em any} \qas (section~\ref{sec:enc}), in
  striking contrast to the situation for classical information, where
  common authentication schemes leave the message completely
  intelligible.  It therefore follows that any authentication protocol
  for an $m$-qubit message must use nearly $2m$ bits of classical key,
  enough to encrypt the message.  The protocol we present approaches
  this bound asymptotically.

\item[{\rm $\bullet$ Impossibility of digitally signing quantum
    states}]{~}
  
  Since authentication requires encryption, it is impossible to create
  digital signature schemes for quantum messages: any protocol which
  allows one recipient to read a message also allows him or her to
  modify it without risk of detection, and therefore all potential
  recipients of an authenticated message must be trustworthy
  (section~\ref{sec:qs}).  This conclusion holds true even if we
  require only computationally secure digital signatures.  Note that
  this does not in any way preclude the possibility of signing {\em
    classical} messages with or without quantum states \cite{GC01}.

\end{description}

Why should we prefer a scheme with classical keys to a scheme with
entangled quantum keys?  The task of authenticating quantum data is
only useful in a scenario where quantum information can be reliably
stored, manipulated, and transmitted over communication lines, so it
would not be unreasonable to assume quantum keys.  However, many
manipulations are easier with classical keys.  Certainly, the
technology for storing and manipulating them is already available, but
there are additional advantages.  Consider, for example, public key
cryptography; it is possible to sign and encrypt classical key bits
with public key systems, but signing a general quantum state is
impossible. Thus, quantum keys would be unsuitable for an asymmetric
quantum authentication scheme such as the one we describe in
section~\ref{PKQA}.
%
% On the other hand, any scheme using $k$ bits of shared classical key
% can be converted in an obvious way to one involving a quantum key
% consisting of $k$ shared EPR pairs, and (although it requires further
% analysis), in such a situation there is potential for adding extra EPR
% pairs to the key, and sacrificing a randomly chosen subset or
% otherwise performing a local protocol, with classical discussion, to
% check for compromise of the key, before using it for authentication.

\section{Preliminaries}
\label{sec:prelim}

\subsection{Classical Authentication}
In the classical setting, an authentication scheme is defined by a pair of 
functions $A:{\cal K}  \times M \fto C$ and
$B:{\cal K}  \times C \fto M\times \{ \mbox{valid} , \mbox{invalid} \}$
such that for any message $\mu\in M$ and key $k\in{\cal K}$ we have {\em completeness}
$$B_k(A_k(\mu))= \langle \mu, \mbox{valid} \rangle$$
and that for any opponent algorithm $O$, we have {\em soundness}
$$\prob{ B_k(O(A_k(\mu))) \in \{ \langle \mu, \mbox{valid} \rangle \} \cup
  \{ \langle \mu', \mbox{invalid} \rangle | \mu'\in M \} } \geq
1-2^{-\Omega(t)}$$
where $t= \lg \#C - \lg \#M$ is the security
parameter creating the tradeoff between the expansion of the messages
and the security level.  Note that we only consider
information-theoretically secure schemes, not schemes that are based
on computational assumptions.

Wegman and Carter \cite{carweg79} introduced several constructions for
such schemes; their most efficient uses keys of size only $4(t +\lg
\lg m) \lg m $ and achieves security $1-2^{-t+2}$.
This compares rather well to the known lower bound of $t+\lg m - \lg
t$ for such a result \cite{carweg79}. The same work also introduced a
technique to re-use an authentication function several times by using
one-time-pad encryption on the tag, so that an opponent cannot learn
{\em anything} about the particular key being used by \Sa\ and \Ra.
Thus, at a marginal cost of only $t$ secret key bits per
authentication, the confidentiality of the authentication function $h$
is guaranteed and thus may be re-used (a polynomial number of times).

\medskip

For the remainder of this paper, we assume the reader is familiar with
the basic notions and notation of quantum computing. These can be
found in textbooks such as \cite{NC00}. Since we rely heavily on
terminology and techniques from quantum error correction (especially
stabilizer codes), appendix~\ref{sec:qecc} provides a summary of the
relevant notions.

\subsection{Purification and purity testing}

Quantum error-correcting codes ({\sc QECC}s) may be used for
\emph{entanglement purification} (\cite{BBPSSW96}). In this setting,
\A\ and \B\ share some Bell states (say
$\ket{\Phi^+}=\ket{00}+\ket{11}$) which have been corrupted by
transmission through a noisy quantum channel. They want a protocol
which processes these imperfect EPR pairs and produces a smaller
number of higher-quality pairs.  We assume that \A\ and \B\ have
access to an authenticated, public classical channel.  At the end of
the protocol, they either accept or reject based on any
inconsistencies they have observed. As long as \A\ and \B\ have a
noticeable probability of accepting, then conditioned on accepting,
the state they share should have fidelity almost 1 to the pure state
$\ket{\Phi^+}^{\otimes m}$.  Moreover, small amounts of noise in their
initial shared state should not cause failure of the protocol.

Stabilizer codes can be particularly useful for purification because
of the following observation: for any stabilizer code $Q$, if we
measure the syndrome of one half of a set of Bell states $\epr{n}$ and
obtain the result $y$, then the result is the state $\epr{m}$, with
each of its two halves encoded in the coset with syndrome $y$.
(Moreover, in this case the distribution on $y$ is uniform.)  If the
original state is erroneous, \A\ and \B\ will likely find different
syndromes, which will differ by the syndrome associated with the
actual error.

Most purification protocols based on stabilizer codes require
efficient error correction; we measure the syndrome, and
use that information to efficiently
restore the encoded state.  However, one can
imagine a weaker task in which Alice and Bob only want to \emph{test}
their EPR pairs for purity, i.e. they want a guarantee that if their
pairs pass the test, their shared state will
probably be close to $\epr{m}$. In that case, we can use the code for error
detection, not correction, and need only be able to
encode and decode efficiently from the space $Q$.

\subsection{Encryption of Quantum Messages}
A useful ingredient for much recent work in quantum cryptography is
the concept of quantum teleportation, put forward by Bennett et al
\cite{BBCJPW00}. After \Sa\ and \Ra\ have shared a singlet state, \Sa\ 
can later secretly send a single qubit in an arbitrary quantum state
$\rho$ to \Ra\ by measuring her half of the singlet state together
with her state $\rho$ in the Bell basis to get two classical bits
$b_0,b_1$. As a result, \Ra's half of the singlet state will become
one of four possibilities $ \rho' := \sigma_{z}^{b_{0}}
\sigma_{x}^{b_{1}} \rho \sigma_{x}^{b_{1}} \sigma_{z}^{b_{0}}$. If
\Sa\ sends $b_0,b_1$, then \Ra\ can easily recover $\rho$.

Now without the bits $b_0,b_1$, the state $\rho'$ reveals no
information about $\rho$. Thus, one can turn this into an encryption
scheme which uses only a classical key: after \Sa\ and \Ra\ have
secretly shared two classical bits $b_{0},b_{1}$, \Sa\ can later
secretly send a single qubit in an arbitrary quantum state $\rho$ to
\Ra\ by sending him a qubit in state $\rho'$ as above. This is called
a quantum one-time pad (QOTP).  This scheme is optimal
\cite{AMTW00,BR00}: any quantum encryption (with a classical key) must
use 2 bits of key for every transmitted qubit.

\section{Quantum Authentication}\label{sec:qa}

At an intuitive level, a quantum authentication scheme is a keyed
system which allows \A\ to send a state $\rho$ to \B\ with a
guarantee: if \B\ accepts the received state as ``good'', the fidelity
of that state to $\rho$ is almost 1.  Moreover, if the
adversary makes no changes, \B\ should always accept, and the fidelity 
should be exactly 1. 

Of course, this informal definition is impossible to attain. The
adversary might always replace \A's transmitted message with a
completely mixed state. There would nonetheless be a small probability
that \B\ would accept, but even when he did accept, the fidelity of the
received state to \A's initial state would be very low.

The problem here is that we are conditioning on \B's acceptance of
the received state; this causes trouble if the adversary's a priori
chances of cheating are high. A more reasonable definition would
require a tradeoff between \B's chances of accepting, and the
expected fidelity of the received system to \A's initial state
given his acceptance: as \B's chance of accepting increases, so
should the expected fidelity.

It turns out that there is no reason to use both the language of
probability and that of fidelity here: for classical tests, fidelity
and probability of acceptance coincide. With this in mind we first
define what constitutes a quantum authentication scheme, and then
give a definition of security:

\begin{defi}\label{def:qap}
  A \emph{quantum authentication scheme} (\qas) is a pair of
  polynomial time quantum algorithms $A$ and $B$ together with a set
  of \emph{classical} keys ${\cal K}$ such that:
  \begin{itemize}
  \item $A$ takes as input an $m$-qubit message system $M$ and a key
    $k\in {\cal K}$ and outputs a transmitted system $T$ of $m+t$
    qubits.
  \item $B$ takes as input the (possibly altered) transmitted system
    $T'$ and a classical key $k\in {\cal K}$ and outputs two systems: a
    $m$-qubit message state $M$, and a single qubit $V$ which
    indicates acceptance or rejection. The classical basis states of
    $V$ are called $\ket{\acc},\ket{\rej}$ by convention.
  \end{itemize}
  For any fixed key $k$, we denote the corresponding super-operators
  by $A_k$ and $B_k$. 
\end{defi}

Note that \B\ may well have measured the qubit $V$ to see whether or
not the transmission was accepted or rejected. Nonetheless, we think
of $V$ as a qubit rather than a classical bit since it will allow us
to describe the joint state of the two systems $M,V$ with a density
matrix.

%(N.B.: In our protocol, $E_k$ will in fact be an isometry from the
%state space of $M$ to the state space of $T$, i.e. for any pure state
%$\ket{\psi}$, the state $E_k\ket{\psi}$ will also be pure.)

There are two conditions which should be met by a quantum
authentication protocol. On the one hand, in the absence of intervention, the
received state should be the same as the initial state and \B\ should
accept.

On the other hand, we want that when the adversary does intervene,
\B's output systems have high fidelity to the statement ``either \B\
rejects or his received state is the same as that sent by \A''. One
difficulty with this is that it is not clear what is meant by ``the
same state'' when \A's input is a mixed state. It turns out that it is
sufficient to define security in terms of pure states; one can deduce
an appropriate statement about fidelity of mixed states (see
Appendix \ref{sec:pfnewdef}).

Given a pure state $\ket{\psi}\in {\cal H}_M$, consider the following
test on the joint system $M,V$: output a 1 if the first $m$ qubits are
in state $\ket{\psi}$ \emph{or} if the last qubit is in state
$\ket{\rej}$ (otherwise, output a 0). The projectors corresponding to this
measurement are
\begin{eqnarray*}
P_1^{\ket{\psi}} &=& \ket{\psi}\bra{\psi}\otimes I_V\ +\ I_M \otimes
\proj{\rej}\ -\ \proj{\psi}\otimes\proj{\rej} \\
P_0^{\ket{\psi}} &=& (I_M-\proj{\psi})\otimes (\proj{\acc})
\end{eqnarray*}
We want that for all possible input states $\ket{\psi}$ and for all
possible interventions by the adversary, the expected fidelity of
\B's output to the space defined by $P_1^{\ket{\psi}}$ is high. This
is captured in the following definition of security.

\begin{defi}\label{def:secure}
  A $\qas$ is secure with error $\epsilon$ for a state $\ket\psi$ if
it satisfies:
  \begin{description}
  \item \emph{Completeness:} For all keys $k\in{\cal K}$: $B_k
    (A_k(\proj{\psi})) = \proj{\psi}\otimes \proj{\acc}$
  \item \emph{Soundness:} For all super-operators ${\cal O}$, let
    $\rho_{Bob}$ be the state output by \B\ when the adversary's
    intervention\footnote{We make no assumptions on the running time
      of the adversary.} is characterized by ${\cal O}$, that is:
    $$\rho_{Bob} = \mathbb{E}_{k} \Big[B_k( {\cal O}(
    A_k(\ket{\psi}\bra{\psi})))\Big] = \frac{1}{|{\cal K}|}\sum_k B_k(
    {\cal O}( A_k(\ket{\psi}\bra{\psi})))$$
    where
    ``~$\mathbb{E}_{k}$'' means the expectation when $k$ is chosen
    uniformly at random from ${\cal K}$. The $\qas$ has soundness error
    $\epsilon$ for $\ket\psi$ if:
    $$\Tr\paren{P_1^{\ket{\psi}} \rho_{Bob} } \geq 1-\epsilon$$
  \end{description}
  A $\qas$ is secure with error $\epsilon$ if it is secure with error
  $\epsilon$ for all states $\ket\psi$.%~\footnote{By linearity, the
%    soundness condition above can be rewritten: 
%    $\quad \mathbb{E}_{k} \left[\Tr\paren{P_1^{\ket{\psi}}D_k( {\cal O}(
%        E_k(\ket{\psi}\bra{\psi}))) }\right] \geq 1-\epsilon$. Thus,
%    it can interpreted as the expected proximity of $\rho_{Bob}$ to
%    the ``secure'' subspace described by $P_1^{\ket\psi}$.}
\end{defi}

Note that our definition of completeness assumes that the channel
connecting \A\ to \B\ is noiseless in the absence of the adversary's
intervention. This is in fact not a significant problem, as we can
simulate a noiseless channel using standard quantum error correction.
% Interestingly, this is not always necessary: the \qas\ we give in the
% following section can be modified to have built-in error tolerance. We 
% discuss this further in Section \ref{AA}.

\paragraph{Interactive protocols}

In the previous section, we dealt only with non-interactive quantum
authentication schemes, since that is both the most natural notion,
and the one we achieve in this paper. However, there is no reason to
rule out interactive protocols in which \A\ and \B\ at the end
believe they have reliably exhanged a quantum message. The definitions
of completeness and soundness extend naturally to this setting: as
before, \B's final output is a pair of systems $M,V$, where the state
space of $V$ is spanned by $\ket\acc,\ket\rej$. In that case
$\rho_{Bob}$ is \B's density matrix at the end of the protocol,
averaged over all possible choices of shared private key and
executions of the protocol. The soundness error is $\epsilon$, where
$\Tr\paren{P_1^{\ket{\psi}} \rho_{Bob} } \geq 1-\epsilon$.

%%%%%%%%%%%%%%%%%%%%%%%%%%%%%%%%%%%%%%%%%%%%%%%%%%%%%%%%%%%%%%%%%%%%%%%%%%
%%%%%%%%%%%%%%%%%%%%%%%%%%%%%%%%%%%%%%%%%%%%%%%%%%%%%%%%%%%%%%%%%%%%%%%%%%
%\newpage
\section{Purity Testing Codes}
\label{sec:ptc}

An important tool in our proof is the notion of a \emph{purity testing
code}, which is a way for \A\ and \B\ to ensure that they share
(almost) perfect EPR pairs.  We shall concentrate on purity testing
codes based on stabilizer QECCs.

\begin{defi} \label{def:ptc}
A stabilizer purity testing code with error $\epsilon$ is a set of
stabilizer codes $\{Q_k\}$, for $k \in {\cal K}$, such that $\forall
\,E_x \in E$ with $x \neq 0$, $\#\{k | x \in Q_k^\perp - Q_k \} \leq
\epsilon (\#{\cal K})$.
\end{defi}

That is, for any error $x$ in the error group, if $k$ is chosen later
at random, the probability that the code $Q_k$ detects $x$ is at least
$1-\epsilon$.

\begin{defi} \label{def:ptp}
A purity testing protocol with error $\epsilon$ is a superoperator
${\cal T}$ which can be implemented with local operations and
classical communicaiton, and which maps $2n$ qubits (half held by \A\
and half held by \B) to $2m+1$ qubits and satisfies the following
two conditions: 
  \begin{description}
  \item \emph{Completeness:} ${\cal T} (\epr{n}) = \epr{m} \otimes
    \ket{\acc}$
  \item \emph{Soundness:} Let $P$ be the projection on the subspace
    spanned by $\epr{m} \otimes \ket{\acc}$ and $\ket{\psi} \otimes
    \ket{\rej}$ for all $\ket{\psi}$.  Then ${\cal T}$ satisfies the
    soundness condition if for all $\rho$,
    $$\Tr\paren{P {\cal T}(\rho)} \geq 1-\epsilon.$$
  \end{description}
\end{defi}

The obvious way of constructing a purity testing protocol ${\cal T}$
is to start with a purity testing code $\{Q_k\}$.  When Alice and Bob
are given the state $\rho$, Alice chooses a random $k \in {\cal K}$
and tells it to Bob.  They both measure the syndrome of $Q_k$ and
compare.  If the syndromes are the same, they accept and perform the
decoding procedure for $Q_k$; otherwise they reject.

\begin{prop} \label{prop:ptp}
If the purity testing code $\{Q_k\}$ has error $\epsilon$, then ${\cal
T}$ is a purity testing protocol with error $\epsilon$.
\end{prop}

The proof appears in Appendix~\ref{sec:pfptp}.

\subsection{An Efficient Purity Testing Code} \label{sec:barnumcode}

Now we will give an example of a particularly efficient purity testing
code.  We will use the stabilizer techniques of
section~\ref{sec:qecc}, restricting to the case $n=rs$.  We will
construct a set of codes $Q_k$ each encoding $m=(r-1)s$ qubits in $n$
qubits, and show that the $Q_k$ form a purity testing code.  (Note
that the construction below works just as well if instead of qubits,
we use registers with dimension equal to any prime power; see
appendix~\ref{app:barnumproof} for details.)  Using qubits in groups
of $s$ allows us to view our field $GF(2^{2rs})$ as both a
$2r$-dimensional vector space over $GF(2^{s})$ and a $2rs$-dimensional
binary vector space.  We need a symplectic form that is compatible
with this decomposition.  One possibility is
\begin{equation}
B(x,y) := \Tr\ (x y^{2^{rs}}),
\label{eq:form}
\end{equation}
where $\Tr(z) = \sum_{i=0}^{2rs-1} z^{2^i}$ is the standard trace
function, which maps $GF(2^{2rs})$ onto $GF(2)$.

We consider a {\em normal rational curve} in $PG(2r-1,2^s)$ (the
projective geometry whose points are the 1-d subspaces of the
$2r$-dimensional vector space over $GF(2^s)$).  (See, e.g., the
excellent introductory text \cite{Beutelspacher98a}.) Such a curve is
given by: 
\beq 
\Upsilon = \{[1:y:y^2:\cdots:y^{2r-1}],[0:0:0:\cdots:1]\}_{y \in
GF(2^s)}.  
\eeq
(The colon is used to separate the
coordinates of a projective point, indicating that only their
ratio matters.)  
Thus, there are $2^s + 1$ points on the normal
rational curve.

Since each ``point'' of this curve is actually a one-dimensional
subspace over $GF(2^s)$, it can also be considered as an
$s$-dimensional binary subspace $Q_k$ in a vector space of dimension
$2rs = 2n$.  We will show that $Q_k$ is totally isotropic with respect
to the symplectic inner product~(\ref{eq:form}), and encodes $m=n-s$
qubits in $n$ qubits.

\begin{theo}
The set of codes $Q_k$ form a stabilizer purity testing code with
error
\beq
\epsilon = \frac{2r}{2^s+1}\;.   
\eeq
Each code $Q_k$ encodes $m=(r-1)s$ qubits in $n=rs$ qubits.
\end{theo}

Proof of this is in Appendix \ref{app:barnumproof}.

%%%%%%%%%%%%%%%%%%%%%%%%%%%%%%%%%%%%%%%%%%%%%%%%%%%%%%%%%%%%%%%%%%%%%%%%%%
%%%%%%%%%%%%%%%%%%%%%%%%%%%%%%%%%%%%%%%%%%%%%%%%%%%%%%%%%%%%%%%%%%%%%%%%%%
%\newpage
\section{Protocols}
\label{sec:prot}

In this section we describe a secure non-interactive quantum
authentication scheme (Protocol \ref{pro4}) which satisfies the
definition of section \ref{sec:qa}.

In order to prove our scheme secure, we begin with a purity testing
protocol as per Section~\ref{sec:ptc} (summarized as Protocol
\ref{pro1}).  The security of this protocol follows from
Prop.~\ref{prop:ptp}.  We then perform several transformations to the
protocol that strictly preserve its security and goals but which
remove the interaction, replacing it with a shared private key.  We
thus obtain two less interactive intermediate protocols (Protocols
\ref{pro2} and \ref{pro3}) and a final protocol (Protocol \ref{pro4}),
which is completely non-interactive.  The transformations are similar
in flavor to those of Shor and Preskill \cite{SP00}, who use the
technique to obtain a simple proof of the security of a completely
different task, namely the BB84 \cite{BB84} quantum key exchange
scheme.

\begin{figure}[h]
  \begin{center}
  \proto{Purity Testing Based Protocol}{\label{pro1}

    \step{\Sa\ and \Ra\ agree on some stabilizer purity testing code
      $\{Q_k\}$} 

    \step{\Sa\ generates $2n$ qubits in state $\ket{\Phi^{+}}^{\otimes
        n}$.  \Sa\ sends the first half of each $\ket{\Phi^{+}}$
      state to \Ra.}

    \step{\Ra\ announces that he has received the $n$ qubits.}
  
    \step{\Sa\ picks a random $k \in {\cal K}$, and announces it to
      \Ra.}
  
    \step{\Sa\ and \Ra\ measure the syndrome of the stabilizer code
      $Q_k$.  \Sa\ announces her results to \Ra\ who compares them to
      his own results.  If any error is detected, \Ra\ aborts.}
  
    \step{\Sa\ and \Ra\ decode their $n$-qubit words according to
      $Q_k$.  Each is left with $m$ qubits, which together should be
      nearly in state $\ket{\Phi^{+}}^{\otimes m}$. }
  
    \step{\Sa\ uses her half of $\ket{\Phi^{+}}^{\otimes m}$ to
      teleport an arbitrary $m$-qubit state $\rho$ to \Ra.}
  }
  \end{center}
\end{figure}

Following the notation of Section \ref{sec:ptc}, let $P$ be the
projector onto the subspace described by ``either \B\ has aborted or
the joint state held by \A\ and \B\ is $\ket{\Phi^+}^{\otimes
m}$~''. Let $\rho_{AB}$ be the joint density matrix of \A\ and \B's
systems. Then Prop.~\ref{prop:ptp} states that at the end of step~6,
$\Tr(P \rho_{AB})$ is exponentially close to 1 in $n$.  The soundness
of our first authentication protocol follows immediately:

\begin{cor}\label{cor:p1}
  If \A\ and \B\ are connected by an authenticated classical
  channel, then Protocol \ref{pro1} is a secure \emph{interactive}
  quantum authentication protocol, with soundness error exponentially
  small in $n$.
\end{cor}

The proof is straightforward; we give it explicitly in
Appendix~\ref{sec:intpro}.

\begin{figure}[h]
  \proto{Non-interactive authentication}{\label{pro4}

    \step{Preprocessing: \Sa\ and \Ra\ agree on some stabilizer purity
      testing code $\{Q_k\}$ and some private and random binary
      strings $k$, $x$, and $y$.}
  
    \step{\Sa\ q-encrypts $\rho$ as $\tau$ using key $x$.  \Sa\
      encodes $\tau$ according to $Q_k$ for the code $Q_k$ with
      syndrome $y$ to produce $\sigma$.  \Sa\ sends the result to
      \Ra.}
  
    \step{\Ra\ receives the $n$ qubits. Denote the received state by
      $\sigma'$.  \Ra\ measures the syndrome $y^{\prime}$ of the code
      $Q_k$ on his qubits. \Ra\ compares $y$ to $y^{\prime}$, and
      aborts if any error is detected.  \Ra\ decodes his $n$-qubit
      word according to $Q_k$, obtaining $\tau'$. \Ra\ q-decrypts
      $\tau'$ using $x$ and obtains $\rho'$. }  }
\end{figure}

\begin{theo}\label{thm:main}
  When the purity testing code $\{Q_k\}$ has error $\epsilon$, the
  protocol \ref{pro4} is a secure quantum authentication scheme with
  key length $O(n + \log_2 (\#{\cal K}))$ and soundness error $\epsilon$.
  In particular, for the purity testing code described in
  Section~\ref{sec:barnumcode}, the authentication scheme has key
  length $2m + s + \log_2 (2^s+1) \leq 2n+1$ and soundness error
  $2n/[s(2^s+1)]$, where $m$ is the length of the message in qubits,
  $s$ is the security parameter, and \A\ sends a total of $n = m+s$
  qubits.
\end{theo}

\proof\ From Corollary \ref{cor:p1} we have that Protocol \ref{pro1} is
a secure interactive authentication protocol. We show that Protocol
\ref{pro4} is equivalent to Protocol \ref{pro1}, in the sense that any
attack on Protocol \ref{pro4} implies an equally succesful attack on
Protocol \ref{pro1}. To do so, we proceed by a series of reductions;
the details appear in Appendix~\ref{sec:intpro}.

\subsection{Public Key Quantum Authentication}\label{PKQA}
Unlike its classical counterpart, quantum information
can be authenticated in a public key setting but not
in a way that can be demonstrated to a judge.  In
section~\ref{sec:enc}, we show the impossibility of a digital signature
scheme for quantum information; here, we instead introduce the
notion of public key quantum authentication.

Let $E_{b}$, $D_{b}$ be \Ra's public and private keyed algorithms to a
PKC resistant to quantum computers' attacks.  Let $S_{a}$, $V_{a}$ be
\Sa's private and public keyed algorithms to a digital signature
scheme resistant to quantum computers' attacks.  These may be either
be protocols which are secure with respect to a computational
assumption~\cite{OTU00} or with unconditional security~\cite{GC01}.
To perform authentication, \Sa\ picks secret and random binary strings
$k$, $x$, and $y$, and uses them as keys to q-authenticate $\rho$ as
$\rho'$.  \Sa\ encrypts and signs the key as $\sigma := S_{a}( E_{b}
(k|x|y) )$.  \Sa\ sends $\left\langle \rho',\sigma \right\rangle$ to
\Ra.
To verify a state, \Ra\ verifies \Sa's signature on $\sigma$ using
$V_{a}$ and then discovers the key $k$, $x$ and $y$ using his private
decryption function $D_{b}$.  \Ra\ checks that $\rho'$ is a valid
q-authenticated message according to key $k$, $x$, $y$, and recovers
$\rho$.

\comment{%%%%%%%%% COMMENT %%%%%%%%%%%%
\subsection{Generalized approach}

Our final scheme of Protocol \ref{pro4} may be simplified even more:
Instead of encrypting message $\rho$ and then encoding it with a
secret syndrome of the code $Q_k$, one may encode $\rho$ according to
the standard syndrome $(y=00\ldots0)$ for the code $Q_k$, and then
encrypt the resulting codeword.  This is true because the full Pauli
group $E$ can be partitioned into cosets of $Q_k^\perp$; different
cosets correspond to different syndromes $y$.  In addition, encrypting
a message which is later encoded using $Q_k$ corresponds to choosing
an element of $Q_k^\perp/Q_k$.  Therefore, the two procedures together
correspond to choosing a random coset of $Q_k$ in the $n$-qubit Pauli
group $E$.  On the other hand, encrypting an $n$-qubit message
requires choosing a completely random Pauli element, which can be done
by first choosing a random coset of $Q_k$ and then a random
representative of the coset.  The additional random choice does not
affect the protocol at all: elements of $Q_k$ act as the identity on
any encoded state, and therefore two elements of the same coset of
$Q_k$ will produce the same final state.

The resulting protocol becomes even more natural: \Sa\ encodes $\rho$
as $\rho'$ using a code $Q_k$ (with $k$ coming from the shared secret
key), and applies QOTP to the whole resulting string. \Ra\ decrypts by
inverting each step.  The price for this aesthetic improvment is a
slight increase in key length from $2m + 2s + 1$ to $2m + 3s + 1$.
}%%%%%%%%% COMMENT %%%%%%%%%%%%

\section{Good Authentication Implies Good Encryption}
\label{sec:enc}

One notable feature of any protocol derived using
Theorem~\ref{thm:main} is that the information being authenticated is
also completely encrypted.  For classical information, authentication
and encryption can be considered completely separately, but in this
section we will show that quantum information is different.  While
quantum states can be encrypted without any form of authentication,
the converse is not true: any scheme which guarantees authenticity
must also encrypt the quantum state almost perfectly.

To show this, let us consider any fixed authentication scheme. Denote
by $\rho_{\ket\psi}$ the density matrix transmitted in this scheme
when Alice's input is $\ket \psi$. Let $\rho_{\ket\psi}^{(k)}$ denote
the density matrix for key $k$.

\begin{defi}
  An encryption scheme with error $\epsilon$ for quantum states hides
  information so that if $\rho_0$ and $\rho_1$ are any two distinct
  encrypted states, then the trace distance $D(\rho_0,\rho_1)=\frac 1
  2 \Tr\abs{\rho_0-\rho_1}\leq\epsilon$.
\end{defi}

We claim that any good \qas\ must necessarily also be a good
encryption scheme.  That is:

\begin{theo}[Main Lower Bound]\label{thm:lowerbd}
  A \qas\ with error $\epsilon$ is an encryption scheme with error at
  most $4\epsilon^{1/6}$.
\end{theo}

\begin{cor}\label{cor:keylen}
  A \qas\ with error $\epsilon$ requires at least $2m
  (1-poly(\epsilon))$ classical key bits.
\end{cor}

We prove this corollary in Appendix \ref{sec:lowerbdpfs}. For now, we
concentrate on the Theorem \ref{thm:lowerbd}.

The intuition behind the proof of this main theorem is that measurement
disturbs quantum states, so if the adversary can learn information
about the state, she can change the state.  More precisely, if the
adversary can distinguish between two states $\ket 0$ and $\ket 1$,
she can change the state $\ket{0} + \ket{1}$ to $\ket{0} - \ket{1}$.
An extreme version of this situation is contained in the following
proposition:

\begin{prop}
  Suppose that there are two states $\ket {0},\ket{1}$ whose
  corresponding density matrices $\rho_ {\ket{0}},\rho_ {\ket{1}}$ are
  perfectly distinguishable. Then the scheme is not an
  $\epsilon$-secure \qas\ for any $\epsilon<1$.
\end{prop}

\begin{proof}
  Since $\rho_ {\ket{0}}$, $\rho_ {\ket{1}}$ can be distinguished,
  they must have orthogonal support, say on subspaces $V_0,V_1$. So
  consider an adversary who applies a phaseshift of $-1$ conditioned
  on being in $V_1$. Then for all $k$, $\rho_{\ket 0 + \ket 1}^{(k)}$
  becomes $\rho_{\ket 0 - \ket 1}^{(k)}$. Thus, Bob will decode the
  (orthogonal) state $\ket 0 - \ket 1$.
\end{proof}

However, in general, the adversary cannot exactly distinguish two
states, so we must allow some probability of failure.  Note that it is
sufficient in general to consider two encoded pure states, since any
two mixed states can be written as ensembles of pure states, and the
mixed states are distinguishable only if some pair of pure states are.
Furthermore, we might as well let the two pure states be orthogonal,
since if two nonorthogonal states $\ket{\psi_0}$ and $\ket{\psi_1}$
are distinguishable, two basis states $\ket 0$ and $\ket 1$ for the
space spanned by $\ket{\psi_0}$ and $\ket{\psi_1}$ are at least as
distinguishable.

Given the space limitations of this abstract, we outline the proof with
a sequence of lemmas, whose proofs are contained in Appendix
\ref{sec:lowerbdpfs}.

We first consider the case when $\ket 0$ and $\ket 1$ can {\em almost}
perfectly be distinguished. In that case, the adversary can change
$\ket{0}+\ket{1}$ to $\ket{0} - \ket{1}$ with high (but not perfect)
fidelity (stated formally in \lemref{almost-dist}).  When $\ket 0$ and
$\ket 1$ are more similar, we first magnify the difference between
them by repeatedly encoding the same state in multiple copies of the
authentication scheme, then apply the above argument.

\begin{lem}\label{lem:almost-dist}
  Suppose that there are two states $\ket {0},\ket{1}$ such that
  $D(\rho_ {\ket 0},\rho_ {\ket 1}) \geq 1-\eta$. Then the scheme is
  not $\epsilon$-secure for $\ket\psi = \ket 0 + \ket 1$ for any
  $\epsilon< 1-2\eta$.
\end{lem}

When two states can be distinguished, but only just barely, the above
lemma is not sufficient.  Instead, we must magnify the
distinguishability of the states $\ket 0$ and $\ket 1$ by repeating
them by considering the tensor product of many copies of the same
state. The probability of distinguishing then goes to 1 exponentially
fast in the number of copies:

\begin{lem}\label{lem:dist}
  Let $\rho_0,\rho_1$ be density matrices with $D(\rho_0,\rho_1)=
  \delta$.  Then $D(\rho_0^{\otimes t},\rho_1^{\otimes t}) \geq 1 -
  2 \exp(-t\delta^2/2)$.
\end{lem}

We create these repeated states by encoding them in an iterated \qas\
consisting of $t$ copies of the original \qas\ (with independent values 
of the key for each copy).

\begin{lem}\label{lem:iterate}
  Suppose we iterate the scheme $t$ times. Let $\ket \psi =
  \frac{1}{\sqrt{2}} (\ket{000...0}+\ket{111...1})$. If $(A,B,{\cal K})$ is an
  $\epsilon$-secure \qas, then the iterated scheme is
  $10t^3\epsilon$-secure for the state $\ket\psi$. 
\end{lem}

Note that the proof of this lemma goes through the following crucial
claim, which follows from a simple hybrid argument.

\begin{claim}[Product states]\label{claim:prod}
  The iterated scheme is $t\epsilon$-secure for any product state.
\end{claim}

Putting the various lemmas together, we find that, given two states
$\ket 0$ and $\ket 1$ which are slightly distinguishable by the
adversary, so $D (\rho_0, \rho_1) \geq \delta$, then in the iterated
scheme, $\ket{000...0}$ and $\ket{111...1}$ are more distinguishable:
$D(\rho_{\ket{000...0}}, \rho_{\ket{111...1}}) \geq 1-\eta$, where
$\eta \leq 2\exp(-t\delta^2/2)$.  Since the iterated scheme is
$10t^3\epsilon$-secure for the state
$\ket\psi=\hfsq(\ket{000...0}+\ket{111...1})$, then by the first
lemma,
$$10t^3\epsilon > 1 - 2\eta \geq 1 -4 \exp(-t\delta^2/2)$$
Choosing $t = 1/\sqrt[3]{20\epsilon}$, we get $\delta \leq 4\epsilon^{1/6}$.

\section{Quantum Signatures}
\label{sec:qs}

One consequence of the previous theorem is that digitally signing
quantum messages is impossible. One can imagine more than one way of
defining this task, but any reasonable definition must allow a
recipient---who should not be able to alter signed messages---to learn
something about the contents of the message. However, this is
precisely what is forbidden by the previous theorem: in an
information-theoretic setting, any adversary who can gain a
non-trivial amount of information must be able to modify the
authenticated state with non-negligible success. 

If we consider computationally secure schemes, a somewhat narrower
definition of digitally signing quantum states remains impossible to
realize.  If we assume a quantum digital signature protocol should
allow any recipient to efficiently extract the original message, then
a simple argument shows that he can also efficiently change it without
being detected, contradicting the security of the scheme.  Namely:
Assume that there is transformation $U$ with a small circuit which
extracts the original message $\rho$, leaving auxiliary state
$\ket\phi$ (which may not all be held by Bob).  In order to preserve
any entanglement between $\rho$ and a reference system, the auxiliary
state $\ket\phi$ must be independent of $\rho$.  Therefore, Bob can
replace $\rho$ with any other state $\rho'$ and then perform
$U^\dagger$ on $\rho'$ and his portion of $\ket\phi$, producing a
valid signature for $\rho'$.  This is an efficient procedure: the
circuit for $U^\dagger$ is just the circuit for $U$ executed
backwards.

Note that we have actually shown a somewhat stronger result: it is not
possible, even when the sender is known to be honest, to authenticate
a quantum message to a group of receivers (some of whom may be
dishonest).  This presentation also makes some limitations of our
proof clear.  For instance, the proof does not apply if the sender
knows the identity of the quantum state he is signing, nor does it
apply to signing classical messages.

\comment{%%%%%%%%% COMMENT %%%%%%%%%%%%
\section{Authentication, Encryption, and Quantum Signatures}
\label{sec:enc}

One notable feature of any protocol derived using
Theorem~\ref{thm:main} is that the information being authenticated is
also completely encrypted.  For classical information, authentication
and encryption can be considered completely separately, but in this
section we will show that quantum information is different.  While
quantum states can be encrypted without any form of authentication,
the converse is not true: any scheme which guarantees authenticity
must also encrypt the quantum state almost perfectly.

To show this, let us consider any fixed authentication scheme. Denote
by $\rho_{\ket\psi}$ the density matrix transmitted in this scheme
when Alice's input is $\ket \psi$. Let $\rho_{\ket\psi}^{(k)}$ denote
the density matrix for key $k$.

\begin{defi}
  An encryption scheme with error $\epsilon$ for quantum states hides
  information so that if $\rho_0$ and $\rho_1$ are any two distinct
  encrypted states, then the trace distance $D(\rho_0,\rho_1)=\frac 1
  2 \Tr\abs{\rho_0-\rho_1}\leq\epsilon$.
\end{defi}

We claim that any good \qas\ must necessarily also be a good
encryption scheme.  That is:

\begin{theo}[Main Lower Bound]\label{thm:lowerbd}
  A \qas\ with error $\epsilon$ is an encryption scheme with error at
  most $4\epsilon^{1/6}$.
\end{theo}

\begin{cor}\label{cor:keylen}
  A \qas\ with error $\epsilon$ requires at least $2m
  (1-poly(\epsilon))$ classical key bits.
\end{cor}

The theorem and corollary are proved in Appendix \ref{sec:lowerbdpfs}.
The intuition behind the proof of the theorem is that measurement
disturbs quantum states, so if the adversary can learn information
about the state, she can change the state.  More precisely, if the
adversary can distinguish between two states $\ket 0$ and $\ket 1$,
she can change the state $\ket{0} + \ket{1}$ to $\ket{0} - \ket{1}$.

One consequence of the previous theorem is that digitally signing
quantum messages is impossible. One can imagine more than one way of
defining this task, but any reasonable definition must allow a
recipient---who should not be able to alter signed messages---to learn
something about the contents of the message. However, this is
precisely what is forbidden by the previous theorem: in an
information-theoretic setting, any adversary who can gain a
non-trivial amount of information must be able to modify the
authenticated state with non-negligible success. 

If we consider computationally secure schemes, a somewhat narrower
definition of digitally signing quantum states remains impossible to
realize.  If we assume a quantum digital signature protocol should
allow any recipient to efficiently extract the original message, then
a simple argument shows that he can also efficiently change it without
being detected, contradicting the security of the scheme.  Namely:
Assume that there is transformation $U$ with a small circuit which
extracts the original message $\rho$, leaving auxiliary state
$\ket\phi$ (which may not all be held by Bob).  In order to preserve
any entanglement between $\rho$ and a reference system, the auxiliary
state $\ket\phi$ must be independent of $\rho$.  Therefore, Bob can
replace $\rho$ with any other state $\rho'$ and then perform
$U^\dagger$ on $\rho'$ and his portion of $\ket\phi$, producing a
valid signature for $\rho'$.  This is an efficient procedure: the
circuit for $U^\dagger$ is just the circuit for $U$ executed
backwards.

Note that we have actually shown a somewhat stronger result: it is not
possible, even when the sender is known to be honest, to authenticate
a quantum message to a group of receivers (some of whom may be
dishonest).  This presentation also makes some limitations of our
proof clear.  For instance, the proof does not apply if the sender
knows the identity of the quantum state she is signing.
}%%%%%%%%% COMMENT %%%%%%%%%%%%

\section{Discussion and Conclusion}

An interesting feature of our scheme: if the transmission quantum
channel is not error free, we can modify our scheme to take advantage
of the error-correction capability of the quantum code.  More precisely,
if \Ra\ rejects only when the number of observed errors is too large
then error correction will fix natural noise or tampering of small
amplitude.

% Moreover, \qas\ can be combined nicely with teleportation so that the
% original state may never be destroyed by an opponent: If \Sa\ 
% succesfully authenticates half of an EPR pair to \Ra\ using the scheme
% with classical keys, she may now teleport any state $\rho$ to \Ra in
% such a way that the opponent must jam \Sa's broadcast of classical
% information to prevent $\rho$ from reaching \Ra. If at any time in the
% future \Sa\ can reliably complete the teleportation with \Ra, then he
% indeed received $\rho$.

We have examined various aspects of the problem of authenticating
quantum messages.  We have shown the security of a large class of
private-key quantum authentication schemes, and presented a particular
highly efficient scheme from that class.  One feature of the scheme is
that it completely encrypts the message, and we show that this is a
necessary feature of any quantum message authentication code: if any
observer can learn a substantial amount of information about the
authenticated state, that observer also has a good chance of
successfully changing the state without being detected.  We have also
studied authentication of quantum states in a public key context, and
shown that while authentication is possible with public keys,
digitally signing quantum states is never possible, even when only
computational security is required.

The necessity for encryption is rather surprising, given that
classical messages can be authenticated without encrypting them.  The
difference can be understood as a complementarity feature of quantum
mechanics: authenticating a message in one basis requires encrypting
it in the complementary Fourier-transformed basis.  This is
essentially another realization of the principle that measuring data
in one basis disturbs it in any complementary basis.  For classical
messages, therefore, encryption is not required: only one basis is
relevant.  In contrast, for quantum messages, we require
authentication in all bases and therefore we must also require
encryption in all bases.

%   An interesting open question is whether
% schemes similar to those used here---for example, the efficient
% purity-testing code alone--suffice to authenticate classical messages,
% via quantum means, more efficiently than classical methods.

Note that purity-testing codes have many applications beyond \qas.
For instance, the efficient purity-testing code of
section~\ref{sec:ptc} can be used to create a correspondingly
efficient QKD protocol.

\section*{Acknowledgments}

We would like to thank Herbert Bernstein, Aart
 Blokhuis, Hoi Fung Chau, David DiVincenzo, Manny
 Knill, Debbie Leung, Michele Mosca, Eric Rains, 
 and Ronald de Wolf for helpful discussions or
 comments. 

%--------------------------------------
\appendix

\section{Quantum Stabilizer Codes}
\label{sec:qecc}

A quantum error-correcting code (QECC) is a way of encoding quantum
data (say $m$ qubits) into $n$ qubits ($m<n$) such that the encoded
data is protected from errors of small weight: the code is said to
correct $t$ errors if any operator which affects less than $t$ qubits
of the encoding can be corrected without disturbing the encoded state.
Usually the goal in the construction of codes is to maximize this
minimum distance for particular $m,n$. However, in this paper, we use
the theory developed for those purposes to construct families of codes
with a different type of property. For now, we review the necessary
theory on a very general class of codes known as \emph{stabilizer
  codes}.

Our construction is based on a class of QECCs for $q$-dimensional
registers, with $q = p^n$ a prime power (later we will specialize to
the case where $p=2$, so each register consists of $n$ qubits). A
basis for the set of all operators on the $p$-dimensional Hilbert
space is the ``shift/phase'' error basis on $p$-dimensional Hilbert
space, defined via $E_{ab} = X^a Z^b$, where $\bra{i}X\ket{j} =
\delta_{i,j+1}$, $\bra{i}Z \ket{j} = \xi^i \delta_{i,j}\;$, for $\xi =
\exp(2\pi i/p)$ a primitive $p$th root of unity, are the
standard-basis matrix elements of the ``shift by one'' and ``ramp the
phase by one'' operators.  (Here, indices are in $\ze_p$.) This basis
has a simple multiplication rule: $E_{ab} E_{a'b'} = \xi^{a'b}
E_{a+a',b+b'}$.  Thus, $\{ \xi^c E_{ab} \}$ is a group containing a
basis for the whole operator space for one register. If we have $n$
registers, we can simply use the tensor product $E$ of $n$ copies of
this operator group; each element corresponds to a $2n$-dimensional
vector, and the vectors $x = ({\bf a}|{\bf b})$, $y = ({\bf a'}|{\bf
  b'})$ come from commuting operators iff their symplectic inner
product is 0 in $\ze_p$:
\begin{equation}
  E_xE_y=E_yE_x \quad \iff \quad B(x,y) = {\bf a'} \cdot {\bf b} -
{\bf a} \cdot {\bf b'}=0. 
\end{equation}
A \emph{stabilizer code} is a QECC given by an Abelian subgroup $S$ of
$E$, which does not contain any multiples of the identity other than
$I$ itself. $S$ can be described by the set of $2n$-dimensional
vectors $x$ such that $E_x \in S$. This will be a subspace of
$\ze_p^{2n}$. Moreover, it will be \emph{totally isotropic}, i.e.
$B(x,y)=0$ for all $x,y$ in the subspace.
If we take a set of
generators for $S$, we can divide Hilbert space into a set of
equidimensional orthogonal subspaces.  Each such space $T$ consists of
common eigenvectors of all operators of $S$ having a fixed pattern of
eigenvalues, unique to $T$.  The space with all eigenvalues $+1$ is
the ``code space,'' its elements are ``codewords,'' and the orthogonal
spaces are labelled by ``syndromes.''

Note that one can also view $B(\cdot,\cdot)$ as a symplectic form over
$GF(p^{2n})$, by choosing a set of generators for $GF(p^{2n})$ as a
vector space over $\ze_p$.  By choosing different sets of generators
for $GF(p^{2n})$ as a vector space over $\ze_p$, we can get different
symplectic forms $B(\cdot,\cdot)$ over this finite vector space.  By
judicious choice of the generators, one can make $B(\cdot,\cdot)$
correspond to \emph{any} non-degenerate symplectic form over
$GF(p^{2n})$.

\mypar{Undetectable errors}
We can classify errors which lie in $E$ into three categories: The
errors corresponding to elements of $Q$ are not truly errors---they
leave the codewords unchanged.  Errors which fail to commute with some
element of $Q$ move codewords into a subspace orthogonal to the code,
so can be detected by the QECC.  The remaining errors, those which
commute with all elements in $S$ but are not themselves in $S$, are
the undetectable errors of the code.  Thus, if $Q^\perp$ is the space
of vectors $y$ for which $B(x,y)=0$ for all $x \in Q$, the set of
undetectable errors is just $Q^\perp - Q$.

\mypar{Syndromes}
Note that specifying the subgroup $S$ by a set $Q$ of elements of
$GF(p^{2n})$ isn't quite enough: operators differing by a phase
$\xi^c$ correspond to the same field element, but yield different
QECC's in the Hilbert space.  
Given an $s$-dimensional totally
isotropic subspace of $\ze_p^{2n}$, there are $p^s$ possible choices
of phases for the group $S$, which produce $p^s$ different QECCs.
However, all these codes have identical error correction properties.
The corresponding code subspaces are all orthogonal and of the same
dimension $p^{n-s}$.  These codes are known as {\em cosets} of the
code $S$, defined as the standard choice with all phases equal to
$1$.\footnote{Actually, the ``standard'' coset also depends on the
selection of a basis of generators for $S$.}  The choice of phases is
known as the {\em syndrome} (because errors outside $S^\perp$ map the
code into a different coset, and the syndrome therefore gives
information about which error occurred).  Measuring the syndrome
projects a quantum state into one of these codes.

\section{Alternative Security Definition}
\label{sec:pfnewdef}

The definition of security of an authentication scheme given in
Section~\ref{sec:qa} appears at first sight to have a major
shortcoming: it does not tell what happens when \A's input is a mixed
state. Intuitively, this should not be a problem, since one expects
security to extend from pure states to mixed states more or less by
linearity. Indeed, this is the case, but it is not entirely clear what
is \emph{meant} by security when \A's input is a mixed state $\rho$.
One straightforward approach is to add a reference system $R$, and to
assume the joint system of \A\ and $R$ is always pure; then the
requirement is that the final state of \B\ and $R$ should high fidelity
to the initial state.  We could also use the following informal
definition, which we will show is implied by Definition
\ref{def:secure}: as long as \B's probability of acceptance is
significant, then when he accepts, the fidelity of the message state
he outputs to \A's original state should be almost 1.

\begin{prop}\label{thm:newdef}
  Suppose that $(A,B, {\cal K})$ is a $\epsilon$-secure $\qas$. Let
  $\rho$ be the density matrix of \A's input state and let $\rho'$
  be the density matrix output by \B\ \emph{conditioned on accepting
    the transmission as valid}. Then if \B's probability of accepting
  is $p_{acc}$, the fidelity of $\rho$ to $\rho'$ is bounded below.
  For any $\rho$ and any adversary action ${\cal O}$, we have:
  $\quad F(\rho,\rho') \geq % \sqrt
  {1- \frac{\epsilon}{p_{acc}}}$.
\end{prop}

In particular, if $\epsilon$ is negligible and $p_{acc}$ is
non-negligible, then the fidelity of \B's state to \A's input state
will be essentially 1.

To prove this, we first restate Proposition \ref{thm:newdef} more
formally. Let $\rho_{Bob}$ be the state of \A's two output systems
$M,V$ when \A's input is $\rho$.  Denote the projector onto the space
of accepting states by $\Pi$, that is $\Pi = I_M \otimes
\proj{\acc}$.

Using this notation, \B's probability of accepting is $p_{acc} =
\Tr(\Pi\rho_{Bob})$, and the density matrix of the joint system $M,V$
conditioned on acceptance is $\rho_{acc} =
\frac{\Pi\rho_{Bob}\Pi}{\Tr(\Pi\rho_{Bob})} =
\frac{\Pi\rho_{Bob}\Pi}{p_{acc}}$.

Now since $\rho_{acc}$ has been restricted to the cases where \B\
accepts, we can write $\rho_{acc}= \rho' \otimes \proj{\acc}$, where
$\rho'$ is the density matrix of \B's message system conditioned on
his acceptance of the transmission as valid. From the definition of
fidelity, we can see that
$$F(\rho,\rho') = F(\rho\otimes \proj{\acc},
  \quad \rho_{acc})$$

We can now restate the theorem:

\noindent \textbf{Claim} (Proposition \ref{thm:newdef}): $\qquad\displaystyle
F(\rho\otimes \proj{\acc},
\ \rho_{acc})\quad \geq \quad {1 - \frac{\epsilon}{p_{acc}}}$

\begin{proofof}{Theorem \ref{thm:newdef}}
%  \textbf{This proof should go in an appendix!}
  Write $\rho= \sum_i p_i \proj{\psi_i}$ for some orthonormal basis
  $\set{\ket{\psi_i}}$. For each $i$, let $\rho_i$ be \B's output
  when \A\ uses input $\ket{\psi_i}$. We have $\rho_{Bob}=\sum_i p_i
  \rho_i$.
  
  For each $i$, let $P_i = \proj{\psi_i}\otimes \proj{\acc}$ and let $Q_i
  = (I_M-\proj{\psi_i})\otimes \proj{\acc}$ so that $P_i+Q_i=\Pi$.
  
  Now we can write $\rho\otimes \proj{\acc} = \sum_i p_i P_i$, and
  $\rho_{acc} = \sum_i p_i\frac{\Pi \rho_i \Pi}{p_{acc}}$. By the
  concavity of fidelity (Theorem 9.7 of \cite{NC00}), we get
  \begin{equation}\label{eq:fid1}
    F(\rho\otimes \proj{\acc},\ \rho_{acc})\ =\  
    F\paren{\sum_i p_i P_i,\ \sum_i
      \frac{\Pi \rho_i \Pi}{p_{acc}} } \ \geq\ \sum_i p_i
    F\paren{P_i,\ \frac{\Pi \rho_i \Pi}{p_{acc}}}
  \end{equation}
  The formula for fidelity for one-dimensional projectors is simple:
  for a projector $P$ and any density matrix $\sigma$, we have
  $F(P,\sigma) = \sqrt{\Tr(P\sigma)}$. Thus expression (\ref{eq:fid1})
  simplifies to
  $$\sum_i p_i \sqrt{\Tr\paren{P_i\frac{\Pi \rho_i \Pi}{p_{acc}}}}$$
  
  Using the fact that $\Pi P_i \Pi =P_i$, we can further simplify this:
  $$\sum p_i \sqrt\frac{\Tr(P_i \rho_i)}{p_{acc}}$$
  
  Since $\frac{\Tr(P_i \rho_i)}{p_{acc}}$ is always less than 1, we
  can obtain a lower bound by removing the square root sign:
   \begin{equation}\label{eq:fid2}
     F(\rho\otimes \proj{\acc}, \rho_{acc})\ \geq\ %\sqrt
     \frac{\sum_i p_i \Tr(P_i \rho_i)}{p_{acc}}
   \end{equation}

%   Since $\sqrt{\cdot}$ is concave, we can apply Jensen's inequality:
%   \begin{equation}\label{eq:fid2}
%     F(\rho\otimes \proj{\acc}, \rho_{acc}) \geq \sqrt\frac{\sum_i p_i \Tr(P_i
%       \rho_i)}{p_{acc}}
%   \end{equation}
  
  Now the acceptance probability $p_{acc}=\Tr(\Pi\rho_{Bob})$ can be
  written as $\sum_i p_i \Tr(\Pi\rho_i)$. Using the fact that $\Pi =
  P_i + Q_i$ we get that $p_{acc} = \paren{\sum_i p_i
    \Tr\paren{P_i\rho_i}}+\paren{\sum_i p_i \Tr\paren{Q_i\rho_i}}$.

  But by the definition of $\epsilon$-security, we know that for each
  $i$, we have $\Tr\paren{Q_i\rho_i} \leq \epsilon$, and so $p_{acc} \leq
\paren{\sum_i p_i
    \Tr\paren{P_i\rho_i}} + \epsilon$, and so we get $\paren{\sum_i p_i
    \Tr\paren{P_i\rho_i}} \geq p_{acc}- \epsilon$. Applying this
  observation to expression (\ref{eq:fid2}), we get :
  $$F(\rho\otimes \proj{\acc}, \rho_{acc})\ \geq\ %\sqrt
  \frac{p_{acc} - \epsilon}{p_{acc} } \ =\ %\sqrt
  {1-\frac{\epsilon}{p_{acc}}}$$
\end{proofof}

%\newpage
\section{Proof of Proposition \ref{prop:ptp}}
\label{sec:pfptp}

Proposition~\ref{prop:ptp} states that a stabilizer purity testing
code can always be used to produce a purity testing protocol with the
same error $\epsilon$.

\begin{proof}
If \A\ and \B\ are given $n$ EPR pairs, this procedure will always
accept, and the output will always be $\epr{m}$.  Thus, ${\cal T}$
satisfies the completeness condition.

Suppose for the moment that the input state is $(E_x \otimes I)
\epr{n}$, for $E_x \in E$, $x \neq 0$.  Then when $k$ is chosen at
random, there is only probability $\epsilon$ that $x \in Q_k^\perp -
Q_k$.  If $x \notin Q_k^\perp$, then \A\ and \B\ will find different
error syndromes, and therefore reject the state.  If $x \in
Q_k^\perp$, then \A\ and \B\ will accept the state, but if $x \in
Q_k$, then the output state will be $\epr{m}$ anyway.  Thus, the
probability that \A\ and \B\ will accept an incorrect state is at most
$\epsilon$.

To prove the soundness condition, we can use this fact and a technique
of Lo and Chau \cite{LC99}.  The states $(E_x \otimes I) \epr{n}$ form
the Bell basis for the Hilbert space of \A\ and \B.  Suppose a
nonlocal third party first measured the input state $\rho$ in the Bell
basis; call this measurement $B$.  Then the argument of the previous
paragraph would apply to show the soundness condition. In fact, it
would be sufficient if Alice and Bob used the nonlocal measurement
$Q_k \otimes Q_k$ which compares the $Q_k$-syndromes for \A\ and \B\
without measuring them precisely.  This is a submeasurement of the
Bell measurement $B$ --- that is, it gives no additional information
about the state.  Therefore it commutes with $B$, so the sequence $B$
followed by $Q_k \otimes Q_k$ is the same as $Q_k \otimes Q_k$
followed by $B$, which therefore gives probability at least
$1-\epsilon$ of success for general input states $\rho$.  But if the
state after $Q_k \otimes Q_k$ gives, from a Bell measurement,
$\epr{m}$ or $\ket{\rej}$ with probability $1-\epsilon$, then the
state itself must have fidelity $1-\epsilon$ to the projection $P$.
Therefore, the measurement $Q_k \otimes Q_k$ without $B$ satisfies the
soundness condition.
Moreover, \A\ and \B's actual procedure ${\cal T}$ is a refinement
of $Q_k \otimes Q_k$---that is, it gathers strictly more information.
Therefore, it also satisfies the soundness condition, and ${\cal T}$
is a purity testing protocol with error $\epsilon$.

\end{proof}

\section{Analysis of Purity-Testing Code Construction}
\label{app:barnumproof}

It is straightforward to extend the purity testing code defined in
Section~\ref{sec:barnumcode} to arbitrary finite fields $GF(q)$.  To
do so, we work over a global field $GF(q^{2rs})$ and break it down
into both a $2r$-dimensional vector space over $GF(q^s)$ and a
$2rs$-dimensional vector space over $GF(q)$.  We exploit this by
defining our $GF(2)$-valued symplectic form $B$ via a choice of a
$GF(q^s)$-valued symplectic form $C$ on $GF(q^{2rs})$ and a non-null
linear map $L:GF(q^s) \rightarrow GF(q)$, where linearity is defined
by viewing $GF(q^s)$ as an $s$-dimensional vector space over
$GF(q)$. Then 
\beq 
B(x,y) := L(C(x,y))\;.
\eeq 
Bilinearity and alternation of $B$ are obvious.  For fixed $y(x)$, by
$C$'s nondegeneracy there is a $z$ such that $C(z,y)(C(x,z)) \ne 0$.
Considering $\alpha z$ in place of $z$, for all scalars $\alpha \in
GF(2^s)$, and still holding $y(x)$ fixed, shows (by bilinearity of
$C$) that $C(x,y)$ takes all values in $GF(q^s)$ as $x(y)$ is varied;
by non-nullity of $L$, not all of these can map to zero, i.e. $B$ is
nondegenerate.

The definition of the purity testing code $\{Q_k\}$ is then the same
as in the binary case.

\begin{theo}
The set of codes $Q_k$ form a stabilizer purity testing code with
error
\beq
\epsilon = \frac{2r}{q^s+1}\;.   
\eeq
Each code $Q_k$ encodes $m=(r-1)s$ dimension $q$ registers in $n=rs$
registers.
\end{theo}

We must show (a) that $Q_k$ is totally isotropic, and (b) that the
error probability is at most $\epsilon$.

(a)
For $\alpha, \beta \in GF(2^s),$ we have 
\bea
B(\alpha x,\beta y) = L(C(\alpha x, \beta y ))
= L(\alpha \beta C(x,y))\;.
\eea
%\iffalse
%(We have used the fact that $\alpha^q = \alpha$ for any field $GF(q)$.
%In particular, $\alpha^{2^{rs}} = \alpha$ and $\beta^{2^{rs}}=
%\beta$.)  
%\fi
Fix an $x \in Q_k-\{0\}$. Every $y \in Q_k$ may be written
as $\alpha x$, for some $\alpha \in GF(q^s)$.  Now $C(x,x)=0$.  So,
$B(\alpha x, \beta x) = 0$, {\em i.e.,} $Q_k$ is totally isotropic
under the symplectic form $B$.

(b) 
We must find, for an arbitrary error $E_x$ (which can be described via
a $2n$-dimensional $GF(q)$ vector $x$), an upper bound on the number
of $Q_k^{\perp} - Q_k$ it can belong to.  It will be sufficient to
bound the number of $Q_k^\perp$ the error can belong to, since $|Q_k|$
is small compared to $|Q_k^\perp|$ in our context.  $x \in Q_k^\perp$
means $B(x,y) = 0$ for all $y \in Q_k$.  By choice of $s$ linearly
independent $y \in Q_k$ this imposes $s$ linearly independent linear
equations on $x$.  We will show below that if we take any $2r$ codes
$Q_k$ defined by points on $\Upsilon$, and take $s$ independent
vectors from each, the resulting set of $2rs$ vectors is linearly
independent.  Thus if $E_x$ is undetectable in $2r$ such codes, this
imposes the dimension's worth ($2rs$) of linearly independent
equations on $x$.  Consequently, $E_x$ must be detectable in all the
remaining codes, i.e., $E_x$ can satisfy $x \in Q_k^\perp$ for {\em at
most} $2r$ values of $k$, when $Q_k$ are chosen among the $q^s+1$
available $s$-dimensional spaces corresponding to points on
$\Upsilon$.  Thus, the $\{Q_k\}$ form a purity testing code with error
\beq 
\epsilon \le \frac{2r}{q^s+1}\;.  
\eeq

We now show the claimed property of codes defined by $\Upsilon$.  A
set of points in a projective geometry of dimension $d-1$ are said to
be in general position if any $d$ (= dimension of the underlying
vector space, when, as in our case, such exists) of them are linearly
independent.  The points on the normal rational curve $\Upsilon$ are
in general position, and indeed a maximal set of such points.  (To
verify that they are in general position one shows that for any $2r$
points on the curve, the determinant of the matrix of their
coordinates is nonzero; these are Vandermonde determinants.)  That is,
any $2r$ points on $\Upsilon$ are linearly independent.  Each point
$k$ on $\Upsilon$ corresponds to an $s$-dimensional code $Q_k$,
consisting of $2rs$-dimensional vectors.  Let $z$ be any nonzero
element of $Q_k$.  As $\alpha$ ranges over $GF(q^s)$, $\alpha z$
ranges over all vectors in $Q_k$.  Thus, if any vector from $Q_k$ is a
linear combination of vectors from other codes $\{Q_j\}$, than all of
$Q_k$ is also a linear combination of vectors from $\{Q_j\}$, and $k$
is linearly dependent on the points $\{j\}$ of $\Upsilon$.  So if we
take any $2r$ codes $Q_k$, and take $s$ independent vectors from each,
the resulting set of $2rs$ vectors is linearly independent.

%\newpage 
\section{Proof of secure authentication}
\label{sec:intpro}

Corollary~\ref{cor:p1} states that the interactive authentication
protocol~\ref{pro1} is secure.

\begin{proofof}{Corollary \ref{cor:p1}}

  The completeness of the protocol can be seen by inspection: in the
  absence of intervention, \A\ and \B\ will share the Bell states
  $\ket{\Phi^+}^{\otimes m}$ at the end of step~6 and so after the
  teleportation in step~7 \B's output will be exactly the input of \A.
  
  To prove soundness, suppose that \A's input is a pure state
  $\ket\psi$. Intuitively, at the end of step~6, \A\ and \B\ share
  something very close to $\ket{\Phi^+}^{\otimes m}$, and so after the
  teleportation in step~7 either \B's output will be very close to
  \A's input, or he will reject because of interference from the
  adversary.
  
  More formally, after step~6, the joint state $\rho_{AB}$ satisfies
  $\Tr(P \rho_{AB}) \geq 1- \epsilon$. At this point, by assumption
  the only thing that the adversary can do is attempt to jam the
  communication between \A\ and \B. Thus the effect of step~7 will be
  to map the subspace given by $P$ into the subspace given by
  $P_1^{\ket\psi}$. Consequently, at the end of the protocol, \B's
  output density matrix will indeed lie almost completely in the
  subspace defined by $P_1^{\ket\psi}$.

\end{proofof}

Theorem~\ref{thm:main} states that the non-interactive
Protocol~\ref{pro4} is secure.  To prove this, we show that
Protocol~\ref{pro1} is equivalent to \ref{pro4}, by moving through two
intermediate protocols~\ref{pro2} and \ref{pro3}.  We reduce the
security of each protocol to the previous one; since
Protocol~\ref{pro1} is secure by Corollary~\ref{cor:p1}, the theorem
follows.

\begin{figure}[h]
  \proto{Intermediate Protocol I}{\label{pro2}
  
    \step{\Sa\ and \Ra\ agree on some stabilizer purity testing code
      $\{Q_k\}$} 

    \step{\Sa\ generates $2n$ qubits in state $\ket{\Phi^{+}}^{\otimes
      n}$.  \Sa\ picks at random $k \in {\cal K}$, and measures the
      syndrome $y$ of the stabilizer code $Q_k$ on the first half of
      the EPR pairs.  \Sa\ decodes her $n$-qubit word according to
      $Q_k$.  \Sa\ performs the Bell measurement to start
      teleportation with her state $\rho$, using the decoded state as
      if it were half of $\ket{\Phi^{+}}$ pairs, but does not yet
      reveal the measurement results $x$ of the teleportation.  \Sa\
      sends the second half of each EPR pair to \Ra.}
  
    \step{\Ra\ announces that he has received the $n$ qubits. Denote
      the received state by $\sigma'$.}
  
    \step{\Sa\ announces $k$ and the syndrome $y$ of $Q_k$ to \Ra.}
  
    \step{\Ra\ measures the syndrome $y^\prime$ of $Q_k$ on his $n$
      qubits.  \Ra\ compares the syndrome $y^\prime$ to $y$.  If they
      are different, \Ra\ aborts.  \Ra\ decodes his $n$-qubit word
      according to $Q_k$.}
  
    \step{\Sa\ concludes the teleportation by sending the
      teleportation measurement results $x$ from step~2. \Ra\ does his
      part of the teleportation and obtains $\rho'$. } 
  }
\end{figure}

\begin{figure}[h]
  \proto{Intermediate Protocol II}{\label{pro3}
  
    \step{\Sa\ and \Ra\ agree on some stabilizer purity testing code
      $\{Q_k\}$} 

    \step{\Sa\ choses a random $2n$ bit key $x$ and q-encrypts $\rho$
      as $\tau$ using $x$.  \Sa\ picks a random $k \in {\cal K}$ and
      syndrome $s$ for the code $Q_k$ and encodes $\tau$ according to
      $Q_k$.  \Sa\ sends the result to \Ra.}
  
    \step{\Ra\ announces that he has received the $n$ qubits. Denote
      the received state by $\sigma'$.}
  
    \step{\Sa\ announces $k$, $x$, and $y$ to \Ra.}
  
    \step{\Ra\ measures the syndrome $y^\prime$ of the code $Q_k$.
      \Ra\ compares $y$ to $y^{\prime}$, and aborts if they are
      different.  \Ra\ decodes his $n$-qubit word according to $Q_k$,
      obtaining $\tau'$. \Ra\ q-decrypts $\tau'$ using $x$ and
      obtains $\rho'$. }  }
\end{figure}

{\sc Protocol \ref{pro1} $\to$ Protocol \ref{pro2}:} We obtain
protocol \ref{pro2} by observing that in protocol \ref{pro1}, \A\ can
perform all of her operations
% of steps \ref{step:sendEPR}, \ref{step:pickk},
% \ref{step:syndrome}, \ref{step:pure}, and \ref{step:teleport} 
(except for the transmissions) \emph{before} she actually sends anything
to \B, since these actions do not depend on \B's feedback.  This will
not change any of the states transmitted in the protocol or computed
by Bob, and so both completeness and soundness will remain the same.

{\sc Protocol \ref{pro2} $\to$ Protocol \ref{pro3}:} There are two
changes between Protocols \ref{pro2} and \ref{pro3}. First, note that
measuring the first qubit of a state $\ket{\Phi^+}$ and obtaining a
random bit $c_i$ is equivalent to choosing $c_i$ at random and
preparing the pure state $\ket{c_i} \otimes \ket{c_i}$.  Therefore,
instead of preparing the state $\ket{\Phi^+}^{\otimes n}$ and
measuring the syndrome of half of it, \A\ may as well choose the
syndromes $s$ at random and encode both halves of the state
$\ket{\Phi^+}^{\otimes m}$ using the code $Q_k$ and the syndrome $s$.

Second, rather than teleporting her state $\rho$ to \B\ using the EPR
halves which were encoded in $Q_{s_1,s_2}$, \A\ can encrypt $\rho$
using a quantum one-time pad (QOTP) and send it to \B\ directly,
further encoded in $Q_k$. These behaviours are equivalent since either
way, the encoded state is $\sigma_x^{\vec t_1}\sigma_z^{\vec t_2}\rho
\sigma_z^{\vec t_2}\sigma_x^{\vec t_1}$, where $\vec t_1$ and $\vec
t_2$ are random $n$-bit vectors.

{\sc Protocol \ref{pro3} $\to$ Protocol \ref{pro4}:} In Protocol
\ref{pro4}, all the random choices of \A\ are replaced with the bits
taken from a secret random key shared only by her and \B. This
eliminates the need for an authenticated classical channel, and for
any interaction in the protocol. This transformation can only increase
the security of the protocol as it simply removes the adversary's
ability to jam the classical communication.  \proofend

\section{Proofs from Section \ref{sec:enc}}
\label{sec:lowerbdpfs}

\begin{theo}[Main Lower Bound]
  A \qas\ with error $\epsilon$ is an encryption scheme with error at
  most $4\epsilon^{1/6}$.
\end{theo}

To get a sense of the proof, consider the following proposition:

\begin{prop}
  Suppose that there are two states $\ket {0},\ket{1}$ whose
  corresponding density matrices $\rho_ {\ket{0}},\rho_ {\ket{1}}$ are
  perfectly distinguishable. Then the scheme is not an
  $\epsilon$-secure \qas\ for any $\epsilon<1$.
\end{prop}

\begin{proof}
  Since $\rho_ {\ket{0}}$, $\rho_ {\ket{1}}$ can be distinguished,
  they must have orthogonal support, say on subspaces $V_0,V_1$. So
  consider an adversary who applies a phaseshift of $-1$ conditioned
  on being in $V_1$. Then for all $k$, $\rho_{\ket 0 + \ket 1}^{(k)}$
  becomes $\rho_{\ket 0 - \ket 1}^{(k)}$. Thus, Bob will decode the
  (orthogonal) state $\ket 0 - \ket 1$.
\end{proof}

However, in general, the adversary cannot exactly distinguish two
states, so we must allow some probability of failure.  Note that it is
sufficient in general to consider two encoded pure states, since any
two mixed states can be written as ensembles of pure states, and the
mixed states are distinguishable only if some pair of pure states are.
Furthermore, we might as well let the two pure states be orthogonal,
since if two nonorthogonal states $\ket{\psi_0}$ and $\ket{\psi_1}$
are distinguishable, two basis states $\ket 0$ and $\ket 1$ for the
space spanned by $\ket{\psi_0}$ and $\ket{\psi_1}$ are at least as
distinguishable.

We first consider the case when $\ket 0$ and $\ket 1$ can {\em almost}
perfectly be distinguished. In that case, the adversary can change
$\ket{0}+\ket{1}$ to $\ket{0} - \ket{1}$ with high (but not perfect)
fidelity (stated formally in \lemref{almost-dist}).  When $\ket 0$ and
$\ket 1$ are more similar, we first magnify the difference between
them by repeatedly encoding the same state in multiple copies of the
authentication scheme, then apply the above argument.

\begin{lem}\label{lem:almost-dist}
  Suppose that there are two states $\ket {0},\ket{1}$ such that
  $D(\rho_ {\ket 0},\rho_ {\ket 1}) \geq 1-\eta$. Then the scheme is
  not $\epsilon$-secure for $\ket\psi = \ket 0 + \ket 1$ for any
  $\epsilon< 1-2\eta$.
\end{lem}

\begin{proofof}{Lemma \ref{lem:almost-dist}}
  Let $\rho_0 = \rho_{\ket 0}$ and $\rho_1=\rho_{\ket 1}$. Consider
  the Hermitian matrix $\sigma=\rho_0-\rho_1$.  We can diagonalize $\sigma$. Let $V_0$ be the space spanned by 
  eigenvectors with non-negative eigenvalues and let $V_1$ be the
  orthogonal complement. 
  
  Since $1/2\,\Tr|\sigma| \geq 1-\eta$, but $\Tr\,\sigma = 0$, we know
  that $\Tr(V_0\sigma) = -\Tr (V_1\sigma) \geq 1-\eta$. Thus,
  $\Tr(V_0\rho_0) \geq \Tr(V_0\sigma) \geq 1 - \eta$. Similarly, $\Tr
  (V_1\rho_1) \geq -\Tr(V_1\sigma) \geq 1-\eta$.
  
  Consider an adversary who applies a phaseshift of $-1$ conditioned
  on being in $V_1$. Fix a particular key $k$. Let
  $p_0=\Tr\paren{V_0\rho_0^{(k)}}$ and
  $p_1=\Tr\paren{V_1\rho_1^{(k)}}$. We know that the expected values
  of $p_0$ and $p_1$ are both at least $1-\eta$.

  \begin{claim} 
    When the input state is $\hfsq (\ket 0+\ket 1)$, the fidelity of
    Bob's output to the state $\hfsq(\ket 0- \ket 1)\ket \acc$ is at
    least $p_0 + p_1 -1$.
  \end{claim}

  \begin{proof} 
    Consider some reference system $R$ which allows us to purify the
    states $\rho_0^{(k)},\rho_1^{(k)}$ to the states $\ket{\tilde
    0},\ket{\tilde 1}$. Let $\ket{\tilde \psi}$ be the image of $\hfsq
    (\ket {\tilde 0} + \ket{ \tilde 1})$ under the adversary's
    conditional phaseshift.

    We want to show that $\ket{\tilde \psi}$ is close to a correct
    encoding of $\hfsq (\ket 0 - \ket 1)$, i.e. close to 
    $$\hfsq (\ket {\tilde
      0}  - \ket{ \tilde 1}) = \hfsq (V_0 \ket{\tilde 0} + V_1\ket{\tilde 0} 
   - V_0\ket{\tilde 1} - V_1\ket{\tilde 1}).$$
   After the transformation, we obtain
   $$\ket{\tilde \psi} = \hfsq (V_0 \ket{\tilde 0} - V_1\ket{\tilde 0} +
   V_0\ket{\tilde 1} - V_1\ket{\tilde 1}).$$
   Thus,
   \begin{eqnarray*}
     \bra{\tilde \psi} \hfsq (\ket{\tilde 0}-\ket{\tilde 1}) & = &
     \frac{1}{2} \left(
       \bra{\tilde 0} V_0 \ket{\tilde 0} - \bra{\tilde 0} V_1 \ket{\tilde 0} -
       \bra{\tilde 1} V_0 \ket{\tilde 1} + \bra{\tilde 1} V_1 \ket{\tilde 1} \right. \\
     & & \left.
       - \bra{\tilde 0} V_0 \ket{\tilde 1} + \bra{\tilde 1} V_0 \ket{\tilde 0}
       + \bra{\tilde 0} V_1 \ket{\tilde 1} - \bra{\tilde 1} V_1 \ket{\tilde 0}
     \right)\\
     &=&
     \frac{1}{2} \left(
       \Tr(V_0\rho_0^{(k)}) - \Tr(V_1\rho_0^{(k)}) -
       \Tr(V_0\rho_1^{(k)}) +\Tr(V_1\rho_1^{(k)})  \right. \\
     & & \left.
       - \Big[\bra{\tilde 0} V_0 \ket{\tilde 1} - \bra{\tilde 1} V_0 \ket{\tilde 0}\Big]
       + \Big[ \bra{\tilde 0} V_1 \ket{\tilde 1} - \bra{\tilde 1} V_1
       \ket{\tilde 0} \Big]
     \right).
   \end{eqnarray*}
   We can substitute for the first line in terms of $p_0$ and $p_1$,
   which are real.  The second line is purely imaginary.  Thus,
   $$\abs{\bra{\tilde \psi} \hfsq (\ket{\tilde 0}-\ket{\tilde 1})} \geq
   \frac{1}{2} \left[ p_0 - (1-p_0) -(1-p_1) + p_1 \right] =
   p_0 + p_1 - 1.$$
   Bob's decoding can only increase the fidelity of the two states, as
   can discarding the reference system, proving the claim.
  \end{proof}
  
  Thus, for a specific value $k$ of the key, $F(\rho^{(k)}, \hfsq
  (\ket{0}-\ket{1}) \ket \acc) \geq p_0 + p_1 - 1$, where $\rho^{(k)}$
  is the output after the adversary's transformation when the input is
  $\hfsq(\ket 0 + \ket 1)$.  Fidelity is concave, so by Jensen's
  inequality the fidelity of the average density matrix $\rho =
  \frac{1}{\abs{\cal K}} \sum_k \rho^{(k)}$ is at least the average of
  the fidelities for each $k$.  That is,
$$F(\rho,  \hfsq (\ket{0}-\ket{1}) \ket \acc) \geq
\frac{1}{\abs{\cal K}} \sum_k (p_0 + p_1 - 1) \geq 1 - 2\eta.$$
In other words, the adversary can change the state $\hfsq(\ket 0 +
\ket 1)$ with probability at least $1-2\eta$.
\end{proofof}

When two states can be distinguished, but only just barely, the above
lemma is not sufficient.  Instead, we must magnify the
distinguishability of the states $\ket 0$ and $\ket 1$ by repeating
them by considering the tensor product of many copies of the same
state. The probability of distinguishing then goes to 1 exponentially
fast in the number of copies:

\begin{lem}\label{lem:dist}
  Let $\rho_0,\rho_1$ be density matrices with $D(\rho_0,\rho_1)=
  \delta$.  Then $D(\rho_0^{\otimes t},\rho_1^{\otimes t}) \geq 1 -
  2 \exp(-t\delta^2/2)$.
\end{lem}

\begin{proofof}{Lemma \ref{lem:dist}}\an{There should be a reference for this.}
  We can bound $D(\rho_0^{\otimes t},\rho_1^{\otimes t})$ by giving a
  test which distinguishes them very well. We know there exists a
  measurement given by spaces $V_0,V_1$ such that $\Tr(V_0 \rho_0) -
  \Tr(V_0\rho_1) =\delta$. Consider the test which performs this
  measurement on $t$ independent copies of $\rho_0$ or $\rho_1$. The
  test outputs 0 if more than $(\Tr(V_0 \rho_0) + \Tr(V_0\rho_1))/2$
  of the measurements produce 0. 
  
  By the Chernoff bound, the probability that this test will make the
  wrong guess is at most $\exp(-t\delta^2/2)$. Thus,
  $D(\rho_0^{\otimes t},\rho_1^{\otimes t}) \geq 1 -
  2\exp(-t\delta^2/2)$.
\end{proofof}

We create these repeated states by encoding them in an iterated \qas\
consisting of $t$ copies of the original \qas\ (with independent values 
of the key for each copy).

\begin{lem}\label{lem:iterate}
  Suppose we iterate the scheme $t$ times. Let $\ket \psi =
  \frac{1}{\sqrt{2}} (\ket{000...0}+\ket{111...1})$. If $(A,B,{\cal K})$ is an
  $\epsilon$-secure \qas, then the iterated scheme is
  $10t^3\epsilon$-secure for the state $\ket\psi$. 
\end{lem}

Note that the proof of this lemma goes through the following crucial
claim, which follows from a simple hybrid argument.

\begin{claim}[Product states]\label{claim:prod}
  The iterated scheme is $t\epsilon$-secure for any product state.
\end{claim}

\begin{proofof}{Claim \ref{claim:prod}}
  For simplicity we prove the claim for the state $\ket{000...0}$. The 
  same proof works for any product pure state (and in fact for
  separable states in general).
  
  Intuitively, an adversary who modifies the state $\ket{000...0}$
  must change some component of the state. We can formalize this by
  rewriting the projector $P_0^{\ket{000...0}}$ in terms of the
  individual projectors $P_0^{\ket {0}_i}$. 

  For the case $t=2$, Bob accepts only if he finds the verification
  qubits for both schemes in the accept state.
  \begin{eqnarray*}
    P_0^{\ket{00}} &=& 
    (I_{m_1 m_2} - \proj{00})\otimes \proj{\acc_1}\otimes \proj{\acc_2} \\
    &=& 
    \Big( (I_{m_1} - \proj{0})\otimes I_{m_2}+    I_{m_1}\otimes
    (I_{m_2} - \proj{0})
    - (I_{m_1} - \proj{0})\otimes (I_{m_2} - \proj{0}) \Big) \\
    && \otimes \proj{\acc_1} \otimes
    \proj{\acc_2} \\
    &=& P_0^{\ket {0}_1}\otimes \proj{\acc_2} + P_0^{\ket
      {0}_2}\otimes \proj{\acc_1} - P_0^{\ket {0}_1} \otimes P_0^{\ket
      {0}_2}
  \end{eqnarray*}
  Since $P_0^{\ket {0}_1} \otimes P_0^{\ket {0}_2}$ is positive, for
  all $\rho$, we have
  $$\Tr(P_0^{\ket{00}}\rho) \leq \Tr(P_0^{\ket {0}_1}\rho) +
  \Tr(P_0^{\ket {0}_2}\rho) \leq 2\epsilon $$
  Similarly, for larger values of $t$ we have 
  $$\Tr(P_0^{\ket{000...0}}\rho) \leq \sum_{i=1}^t \Tr(P_0^{\ket
    {0}_i}\rho)\leq t\epsilon $$
  Thus the iterated scheme is $t\epsilon$-secure for $\ket{000...0}$
  (and in fact for all product states).
\end{proofof}

\begin{proofof}{Lemma \ref{lem:iterate}}
  Consider the net superoperator due to encoding, decoding, and the
  adversary's intervention, i.e. ${\cal O}_{net} = \frac 1 {|{\cal
      K}|}\sum_k B_k {\cal O}_{adv} A_k$. By introducing an ancilla
  system $R$, we can extend this superoperator to a linear
  transformation on the joint system $M \otimes R \otimes V $ (where
  $M$ is the message system and $V$ is Bob's verifcation qubit).
  For a pure state $\ket\psi$, write its image as 
  $$\ket\psi \ket{\gamma_{\ket\psi}}\ket{\acc} + \ket{\beta_
    {\ket\psi}}\ket {\rej} + \ket{\delta_{\ket\psi}}\ket\acc$$
  where $\ket{\delta_{\ket\psi}}$ is a joint state of $MR$ which is
  orthogonal to the subspace $\ket\psi \otimes R$.
  
  Now consider the family of states $\ket{\psi_i} =
  \ket{\underbrace{000...0}_{i} \underbrace{111...1}_{t-i}}$, and let
  $\ket{\gamma_i}=\ket{\gamma_{\ket{\psi_i}}}$ and $\ket{\delta_i} =
  \ket{\delta_{\ket{\psi_i}}}$.

  \begin{claim}
    For all $i=0,...,t-1$, we have
    $\|\frac 1 2 (\ket{\gamma_{i+1}}-\ket{\gamma_i})\| \leq (1+\sqrt{2})\sqrt{t\epsilon}$ 
  \end{claim}
  \begin{proof}
    Fix $i$. Note that $\ket{\psi_+} =
    \hfsq(\ket{\psi_{i+1}}+\ket{\psi_i})$ is a product state (with
    $H\ket{0}$ in one position), as is $\ket{\psi_-} =
    \hfsq(\ket{\psi_{i+1}}-\ket{\psi_i})$. The image of $\ket{\psi_+}$
    can be written
    \begin{eqnarray*}
      \lefteqn{\hfsq\Big( (\ket{\psi_{i+1}}\ket{\gamma_{i+1}}+
      \ket{\psi_i}\ket{\gamma_i} ) \ket \acc +
      (\ket{\delta_{i+1}}+\ket{\delta_i}) \ket\acc +
      (\ket{\beta_{i+1}}+ \ket{\beta_i})\ket\rej \Big)} \\
      &=& \Big( \ket{\psi_+}\frac {1}{2} (\ket{\gamma_{i+1}}+\ket{\gamma_i}) 
      + \ket{\psi_-} \frac {1}{2} (\ket{\gamma_{i+1}}-\ket{\gamma_i})
      + \hfsq(\ket{\delta_{i+1}}+\ket{\delta_i}) \Big) \ket\acc \\
      &&+
      \hfsq(\ket{\beta_{i+1}}+ \ket{\beta_i})\ket\rej
    \end{eqnarray*}
    Now we know that $\|\ket{\delta_i}\|^2 \leq t\epsilon$ for all
    $i$ (since $\ket{\gamma_i}$ is a product state). Thus, $\|
    \hfsq(\ket{\delta_{i+1}}+\ket{\delta_i})\| \leq
    \sqrt{2t\epsilon}$.  

    Moreover, $\ket{\psi_+}$ is a product state and so we have 
    $$\| \ket{\psi_-} \frac {1}{2} (\ket{\gamma_{i+1}}-\ket{\gamma_i})
    + \hfsq(\ket{\delta_{i+1}}+\ket{\delta_i}) \| \leq
    \sqrt{t\epsilon}$$
    Thus, $\| \ket{\psi_-} \frac {1}{2}
    (\ket{\gamma_{i+1}}-\ket{\gamma_i})\| = \| \frac {1}{2}
    (\ket{\gamma_{i+1}}-\ket{\gamma_i})\| \leq (1+\sqrt{2})\sqrt{t\epsilon}$.
  \end{proof}

  Then by the triangle inequality, 
  $\|\frac 1 2 (\ket{\gamma_t} - \ket{\gamma_0})\| \leq
  (1+\sqrt{2})t \sqrt{t\epsilon}$. Let $\ket{\Psi_\pm} =
  \hfsq(\ket{\psi_{t}}\pm\ket{\psi_0})$. The image of
  $\ket{\Psi_+}=\hfsq(\ket{000...0}+\ket{111...1})$ is:
  \begin{eqnarray*}
    &&\Big( \ket{\Psi_+}\frac {1}{2} (\ket{\gamma_t}+\ket{\gamma_0}) 
    + \ket{\Psi_-} \frac {1}{2} (\ket{\gamma_t}-\ket{\gamma_0})
    + \hfsq(\ket{\delta_t}+\ket{\delta_0}) \Big) \ket\acc \\
    &&+
    \hfsq(\ket{\beta_t}+ \ket{\beta_0})\ket\rej
  \end{eqnarray*}
  Now the trace of this state with $P_0^{\ket{\Psi_+}}$ is the square of
  \begin{eqnarray*}
    \|\ket{\Psi_-} \frac {1}{2} (\ket{\gamma_t}-\ket{\gamma_0}) +
    \hfsq(\ket{\delta_t}+\ket{\delta_0})\| & \leq &  \|\ket{\Psi_-} \frac
    {1}{2} (\ket{\gamma_t}-\ket{\gamma_0})\| +
    \|\hfsq(\ket{\delta_t}+\ket{\delta_0})\| \\
    & \leq & (1+\sqrt{2})t \sqrt{t\epsilon} +  \sqrt{2t\epsilon} \\
    & \leq & \sqrt{10 t^3 \epsilon},
  \end{eqnarray*}
  where in the last line, we have assumed $t\geq2$.  That is, the
  iterated scheme is $10t^3\epsilon$-secure for $\ket{\Psi_+}$.  
\end{proofof}

Putting the various lemmas together, we find that, given two states
$\ket 0$ and $\ket 1$ which are slightly distinguishable by the
adversary, so $D (\rho_0, \rho_1) \geq \delta$, then in the iterated
scheme, $\ket{000...0}$ and $\ket{111...1}$ are more distinguishable:
$D(\rho_{\ket{000...0}}, \rho_{\ket{111...1}}) \geq 1-\eta$, where
$\eta \leq 2\exp(-t\delta^2/2)$.  Since the iterated scheme is
$10t^3\epsilon$-secure for the state
$\ket\psi=\hfsq(\ket{000...0}+\ket{111...1})$, then by the first
lemma,
$$10t^3\epsilon > 1 - 2\eta \geq 1 -4 \exp(-t\delta^2/2)$$
Choosing $t = 1/\sqrt[3]{20\epsilon}$, we get $\delta \leq 4\epsilon^{1/6}$.

\begin{cor}
  A \qas\ with error $\epsilon$ requires at least $2m
  (1-poly(\epsilon))$ classical key bits.
\end{cor}

\begin{proofof}{Corollary \ref{cor:keylen}}
  The argument is similar to the argument that $2m$ bits of key are
  required for perfect encryption.  We show that transmitting the key
  through a channel allows the transmission of almost $2m$ bits of
  information.

  We can consider four subsystems, two held by Alice and two held by
  Bob.  Bob holds both halves of $m$ Bell states (the subsystems $B_1$
  and $B_2$), except that $B_1$ has been encrypted by a key $k$
  (subsystem $K$) held by Alice.  Alice also holds $R$, a purification 
  of the other three systems.
  
  Using superdense coding, Bob's two systems $B_1$ and $B_2$ can encode
  $2m$ classical bits of information.  In order to recover that
  information, Bob needs Alice's key (system $K$).  Since the
  encryption is not perfect, however, Bob may have a small amount of
  information about the encoded state.
  
  Let us imagine that Bob's systems initially encode the classical
  message $000...0$.  Suppose Alice wishes to send Bob the message
  $M$.  Since the encryption is almost perfect, Bob's two density
  matrices $\rho_B(000...0)$ and $\rho_B(M)$ are almost
  indistinguishable.  Therefore, by the argument proving bit
  commitment is impossible, Alice can change the pure state
  corresponding to encrypted $000...0$ to something very close to the
  pure state corresponding to encrypted $M$.  
  
  If Alice now sends $K$ to Bob, he is able to (almost always) decode
  the message $M$.  His failure probability is a polynomial in
  $\epsilon$, so he has received $2m(1-poly(\epsilon))$ bits of
  information, and therefore $K$ must consist of at least
  $2m(1-poly(\epsilon))$ classical bits or half as many qubits.  
  
  In fact, $K$ might as well be classical: Bob's decoding method will
  be to immediately measure $K$, since he is expecting a classical
  key, and therefore Alice might as well have measured $K$ before
  sending it; naturally, this actually means she includes entangled
  qubits in the purification $R$.  We thus restrict $K$ to classical
  bits and prove the corollary.
\end{proofof}

\end{document}